\documentclass[aps,twocolumn,epsf,floats,pre]{revtex4}
\usepackage{graphics,graphicx,epsfig}
\usepackage{amssymb,color}
\usepackage{epsf,epstopdf,wrapfig}

\begin{document}

\title{Statistical mechanics for natural flocks of birds}

\author{William Bialek,$^a$ Andrea Cavagna,$^b$ Irene Giardina,$^b$ Thierry Mora,$^c$ Edmondo Silvestri,$^b$ Massimiliano Viale,$^b$ and Aleksandra M Walczak$^d$}

\affiliation{$^a$Joseph Henry Laboratories of Physics and Lewis--Sigler Institute for Integrative Genomics, Princeton University, Princeton, New Jersey 08544 USA\\
$^b$Istituto Sistemi Complessi (ISC--CNR), Via dei Taurini 19, 00185 Roma, Italy\\
$^b$Dipartimento di Fisica, ``Sapienza'' Universit\'a di Roma,  P.le Aldo Moro 2, 00185 Roma, Italy\\
$^c$Laboratoire de Physique Statistique de lÕ'{\'{E}cole} Normale Sup\'erieure, CNRS and Universities Paris VI and Paris VII, 24 rue Lhomond, 75231 Paris Cedex 05, France,  and\\
$^d$Laboratoire de Physique Th\'eorique de lÕ'\'Ecole Normale Sup\'erieure, CNRS and University Paris VI, 24 rue Lhomond, 75231 Paris Cedex 05, France}

\begin{abstract}
Interactions among neighboring birds in a flock cause an alignment of their flight directions.  We show that the minimally structured (maximum entropy) model consistent with these local correlations correctly predicts the propagation of order throughout entire flocks of starlings, with no free parameters.    These models are mathematically equivalent to the Heisenberg model of magnetism, and define an ``energy'' for each configuration of flight directions in the flock.  Comparing flocks of different densities, the range of interactions that contribute to the energy involves a fixed number of (topological) neighbors, rather than a fixed (metric) spatial range.  Comparing flocks of different sizes, the model correctly accounts for the observed scale invariance of long ranged correlations among the fluctuations in flight direction.
\end{abstract}

\date{\today}

\maketitle

The collective behaviour of large groups of a\-ni\-mals is an imposing natural phenomenon, very hard to cast into a systematic theory \cite{camazine+al_01}. Physicists have long hoped that such collective behaviours in biological systems could be understood in the same way as we understand collective behaviour in physics, where statistical mechanics provides a bridge between microscopic rules and macroscopic phenomena \cite{anderson_72,wilson_79}.  A natural test case for this approach is the emergence of order in a flock of birds:  out of a network of distributed interactions among the individuals, the entire flock spontaneously chooses a unique direction in which to fly \cite{couzin+krause_03}, much as local interactions among individual spins in a ferromagnet lead to a spontaneous magnetization of the system as a whole \cite{magnets}.  Despite detailed development of these ideas \cite{toner+tu_95,viscek+al_95,toner+tu_98,gregoire+chate_04}, there still is  a gap between theory and experiment. Here we show how to bridge this gap, by constructing a maximum entropy model \cite{jaynes_57} based on field data of large flocks of starlings \cite{ballerini+al_08a,cavagna+al_08a,cavagna+al_08b}. We use this framework to show that the effective interactions among birds are local, and that the number of interacting neighbors is independent of flock density, confirming that interactions are ruled by topological rather than metric distance \cite{ballerini+al_08b}.  The statistical mechanics models that we derive in this way provide an essentially complete, parameter free theory for the propagation of directional order  throughout the flock.

\begin{figure}
\includegraphics[width = 0.46 \linewidth]{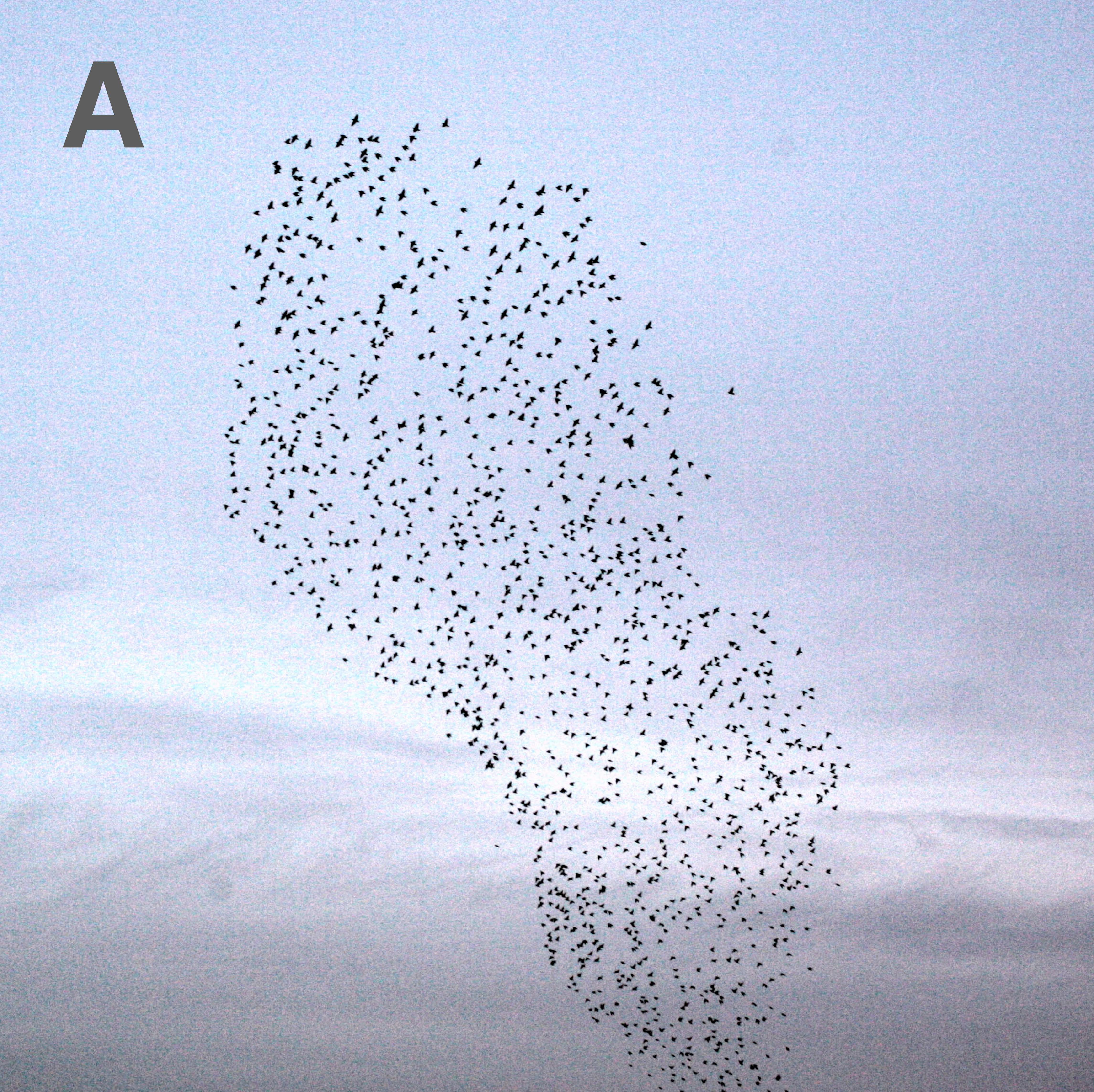}
\includegraphics[width = 0.52 \linewidth]{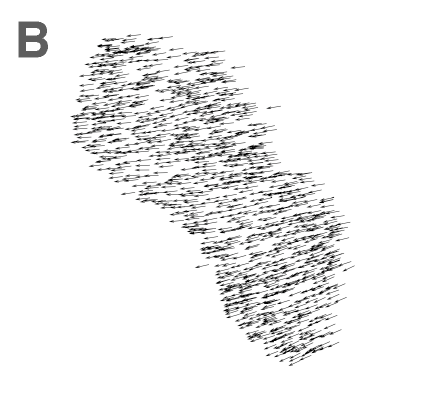}\\
\caption{The raw data.  (A) One snapshot from flocking event 28--10,  $N=1246$ birds (see Table \ref{table} in Methods). (B) Instantaneous vector velocities of all the individuals in this snapshot, normalized as ${\vec s}_{\rm i} = {\vec v}_{\rm i} /| {\vec v}_{\rm i}|$.  
\label{vectors}}
\end{figure}

We consider flocks of European starlings {\em Sturnus vulgaris}, as in Fig \ref{vectors}A.   At any given instant of time, following Refs \cite{ballerini+al_08a,cavagna+al_08a,cavagna+al_08b}, we can attach to each bird $\rm i$ a vector velocity ${\vec v}_{\rm i}$,  and define the normalized velocity ${\vec s}_{\rm i} = {\vec v}_{\rm i} /| {\vec v}_{\rm i}|$ (Fig \ref{vectors}B). On the hypothesis that flocks have statistically stationary states, we can think of all these normalized velocities as being drawn (jointly) from a probability distribution $P(\{{\vec s}_{\rm i}\})$. It is not possible to infer this full distribution directly from experiments, since the space of states specified by $\{{\vec s}_{\rm i}\}$ is too large.
However, what we can measure from field data is the matrix of correlations between the normalized velocities, $C_{\rm ij}= \langle {\vec s}_{\rm i} {\bf \cdot} {\vec s}_{\rm j}\rangle $. There are infinitely many distributions $P(\{  {\vec s}_{\rm i}\})$ that are consistent with the measured correlations, but out of all these distributions, there is {\it one} that has minimal structure: it describes a system that is as random as it can be while still matching the experimental data.  This distribution is the one with maximum entropy \cite{jaynes_57}. 

It should be emphasized that the maximum entropy principle is not a ``modeling assumption;'' rather it is the absence of assumptions.  Any other model that accounts for the observed correlations will have more structure, and hence (explicitly or implicitly) assumes something about the nature of the interactions in the flock beyond what is required to match the data. Of course the fact that the maximum entropy model is minimally structured does not make it correct.  It could be, for example, that individual birds set their flight direction by computing a complicated nonlinear combination of the velocities from multiple neighbors, in which case correlations among pairs of birds would be insufficient to capture the essence of the ordering mechanism. We view the maximum entropy model as a powerful starting point, from which, as we will see, we can generate detailed and testable predictions.

The maximum entropy distribution consistent with the directional correlations $C_{\rm ij}$ is 
\begin{equation}
P(\{  {\vec s}_{\rm i}\}) = {1\over {Z(\{J_{\rm ij}\})}} \exp\left[ {1\over 2}\sum_{{\rm i}=1}^N \sum_{{\rm  j}=1}^N J_{\rm ij}  \; {\vec s}_{\rm i} {\bf \cdot} {\vec s}_{\rm j}\right] ,
\label{maxent1}
\end{equation}
where  $Z(\{J_{\rm ij}\})$ is the appropriate normalization factor, or partition function; the derivation follows Ref \cite{jaynes_57}, as explained in Appendix \ref{App:maxent}.   Notice that there is one parameter $J_{\rm ij}$ corresponding to each measured element $C_{\rm ij}$ of the correlation matrix.    To finish the construction of the model, we have to adjust the values of the $J_{\rm ij}$ to match the experimentally observed $C_{\rm ij}$,
\begin{equation}
\langle {\vec s}_{\rm i} {\bf \cdot} {\vec s}_{\rm j} \rangle_{P}=\langle {\vec s}_{\rm i} {\bf \cdot} {\vec s}_{\rm j} \rangle_{\rm exp}  \ ,
\label{unzu}
\end{equation}
where the symbol $\langle\cdot\rangle_{P}$ indicates an average using distribution $P$ from Eq (\ref{maxent1}), whereas $\langle\cdot\rangle_{\rm exp}$ indicates an average over many experiments.   This matching condition is equivalent to maximizing the likelihood that the model in Eq (\ref{maxent1}) will generate the data from which the correlations were computed.

The probability distribution in Eq (\ref{maxent1}) is mathematically identical to a model that is familiar from the physics of magnets---the Heisenberg model \cite{magnets}---in which a collection of spins ${\vec s}_{\rm i}$ interact so that their energy (or Hamiltonian) is $H(\{ {\vec s}_{\rm i}\}) = - (1/2) \sum_{\rm i,j} J_{\rm ij} \; {\vec s}_{\rm i} {\bf \cdot} {\vec s}_{\rm j}$; Eq (\ref{maxent1}) then describes the thermal equilibrium or Boltzmann distribution at a temperature $k_B T = 1$.   In this context, the constants $J_{\rm ij}$ are the strength of interaction between spins $\rm i$ and $\rm j$, where $J > 0$ means that these elements tend to align.  For many physical systems, once we know the Hamiltonian there is a plausible dynamics that allows the system to relax toward equilbrium, and this is the Langevin dynamics
\begin{equation}
\frac{d{\vec s}_{\rm i}}{dt} = -\frac{\partial H}{\partial \vec s_{\rm i}}+\vec\eta_{\rm i}(t)=\sum_{{\rm j}=1}^N J_{\rm ij} {\vec s}_{\rm j} + \vec\eta_{\rm i}(t) \ ,
\label{langevin}
\end{equation}
where $\vec\eta_{\rm i}(t)$ is an independent white noise ``force'' driving each separate degree of freedom.  Finding trajectories ${\vec s}_{\rm i} (t)$ that solve Eq (\ref{langevin}) produces  samples that are drawn out of the probability distribution in Eq (\ref{maxent1}). The interesting point is that this kind of dynamical model also is well known in biology: the direction of motion of an individual evolves in time according to ``social forces'' reflecting a weighted sum of inputs from neighboring individuals, plus noise \cite{couzin+krause_03}. In this framework, $J_{\rm ij}$ measures the strength of the force that tries to align the velocity of bird $\rm i$ along the direction defined by bird $\rm j$.  We emphasize that this is not an analogy, but a mathematical equivalence.  

In contrast to most networks, the  connectivity in a flock of birds is intrinsically dynamic---birds move and change their neighbors.   Thus, it may not make sense to talk about matrix of correlations $C_{\rm ij}$ or interactions $J_{\rm ij}$ between labelled individuals.  On the other hand, the continuous and rapid change of neighbors induced by motion implies that the interaction $J_{\rm ij}$ between bird $\rm i$ and bird $\rm j$ cannot depend directly on their specific identities, but only on some function of their relative positions.

The simplest form of  interaction that is independent of the birds' identity is one in which each bird interacts with the same strength, $J$, with the same number of neighbors, $n_c$ (or with all birds within the same radius $r_c$; see below).  If the interactions are of this form, then Eq (\ref{maxent1}) simplifies to 
\begin{equation}
P(\{{\vec s}_{\rm i}\}) = {1\over {Z(J,n_c)}} \exp\left[ \frac{J}{2}\sum_{{\rm i}=1}^N \sum_{{\rm j} \in n^{\rm i}_c} {{\vec s}_{\rm i}} {\bf \cdot}  {{\vec s}_{\rm j}} \right]  \ ,
\label{maxent_local}
\end{equation}
where ${\rm j}\in n_c^{\rm i}$ means that bird $\rm j$ belongs to the first $n_c$ nearest neighbors of $\rm i$.  This is, in fact, the maximum entropy model consistent with the average correlation among birds within the neighborhood defined by $n_c$, 
\begin{equation}
C_{\rm int} = {1\over N} \sum_{{\rm i}=1}^N{1\over{n_c}} \sum_{{\rm j} \in n_c^{\rm i}} \langle {{\vec s}_{\rm i}} {\bf \cdot} {{\vec s}_{\rm j}} \rangle
\approx {1\over N} \sum_{{\rm i}=1}^N{1\over{n_c}} \sum_{{\rm j} \in n_c^{\rm i}} {{\vec s}_{\rm i}} {\bf \cdot} {{\vec s}_{\rm j}}  \ .
\label{corre_local}
\end{equation}
Biologically, Eq's  (\ref{maxent_local}) and (\ref{corre_local})  encapsulate the concept that the fundamental correlations are between birds and their directly interacting neighbors; all more distant correlations should be derivable from these. As in the more general problem, finding the values of $J$ and $n_c$ that reproduce the observed correlation $C_{\rm int}$ is the same as maximizing the probability, or likelihood, that model Eq (\ref{maxent_local}) generates the observed configuration of flight directions $\{{\vec s}_{\rm i}\}$ in a single snapshot. If this is correct, a model that appropriately reproduces the fundamental correlations up to the scale $n_c$ must be able to describe correlations on all length scales. 

\begin{figure*}
\includegraphics[width = 0.3 \linewidth]{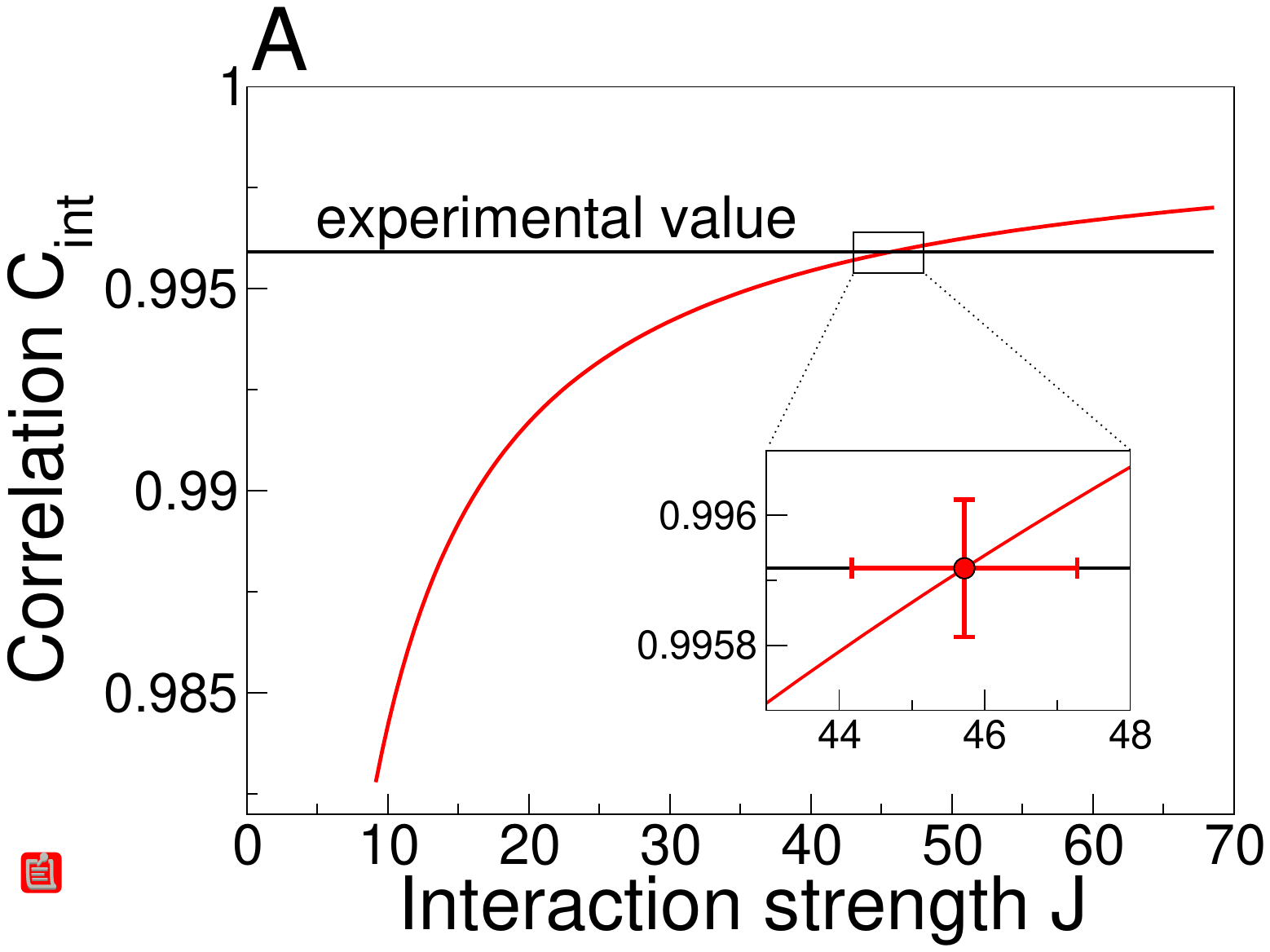} 
\includegraphics[width = 0.3 \linewidth]{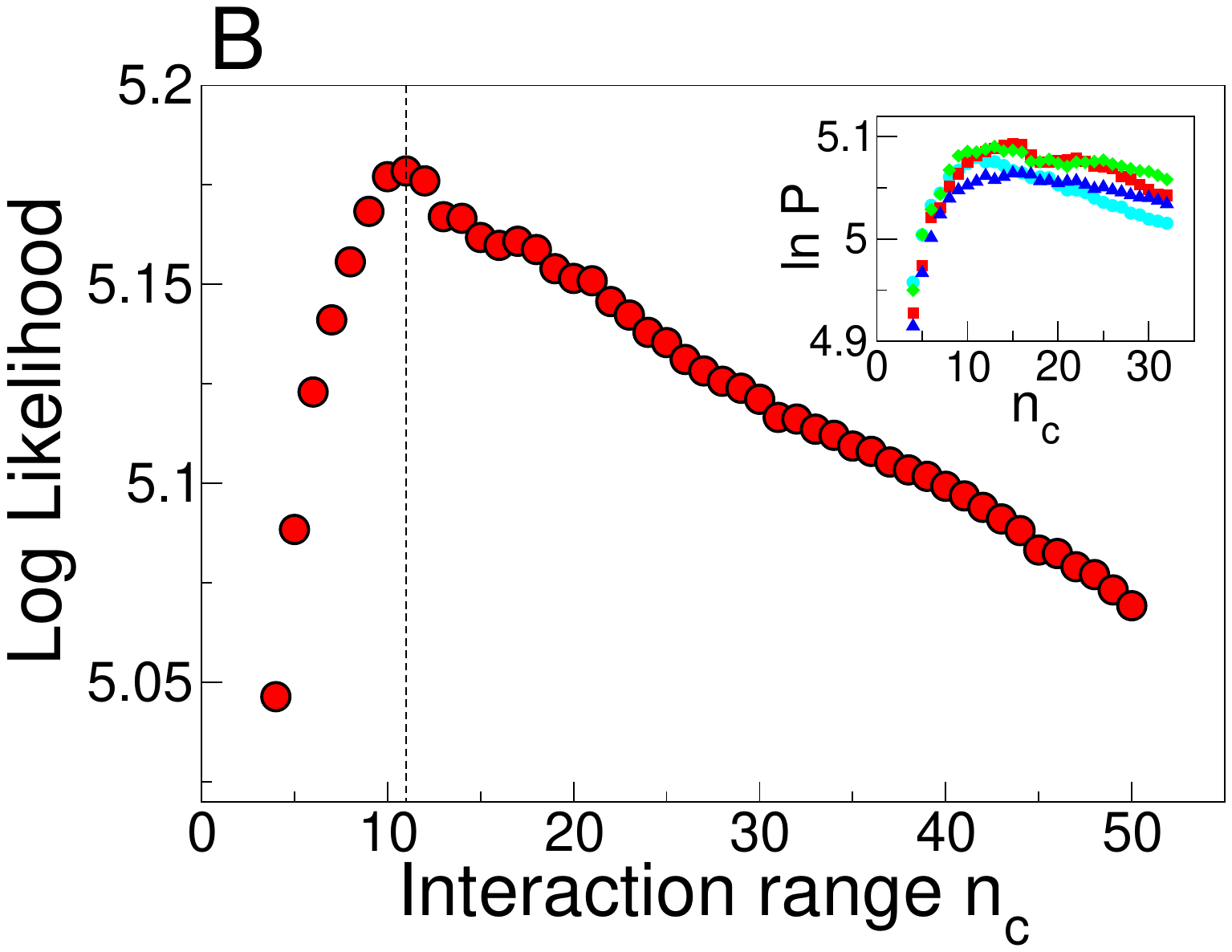}
\includegraphics[width = 0.3 \linewidth]{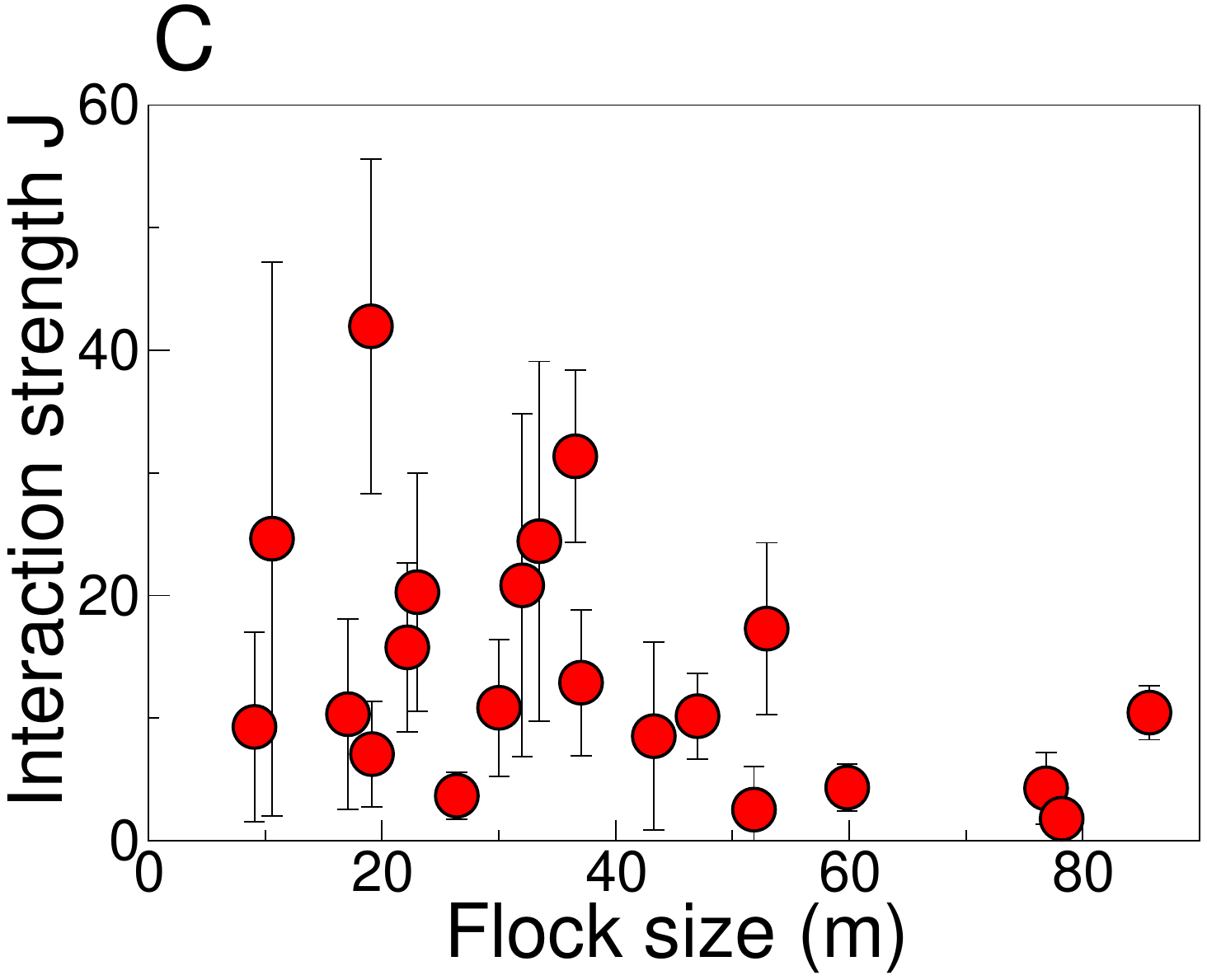}
\includegraphics[width = 0.3 \linewidth]{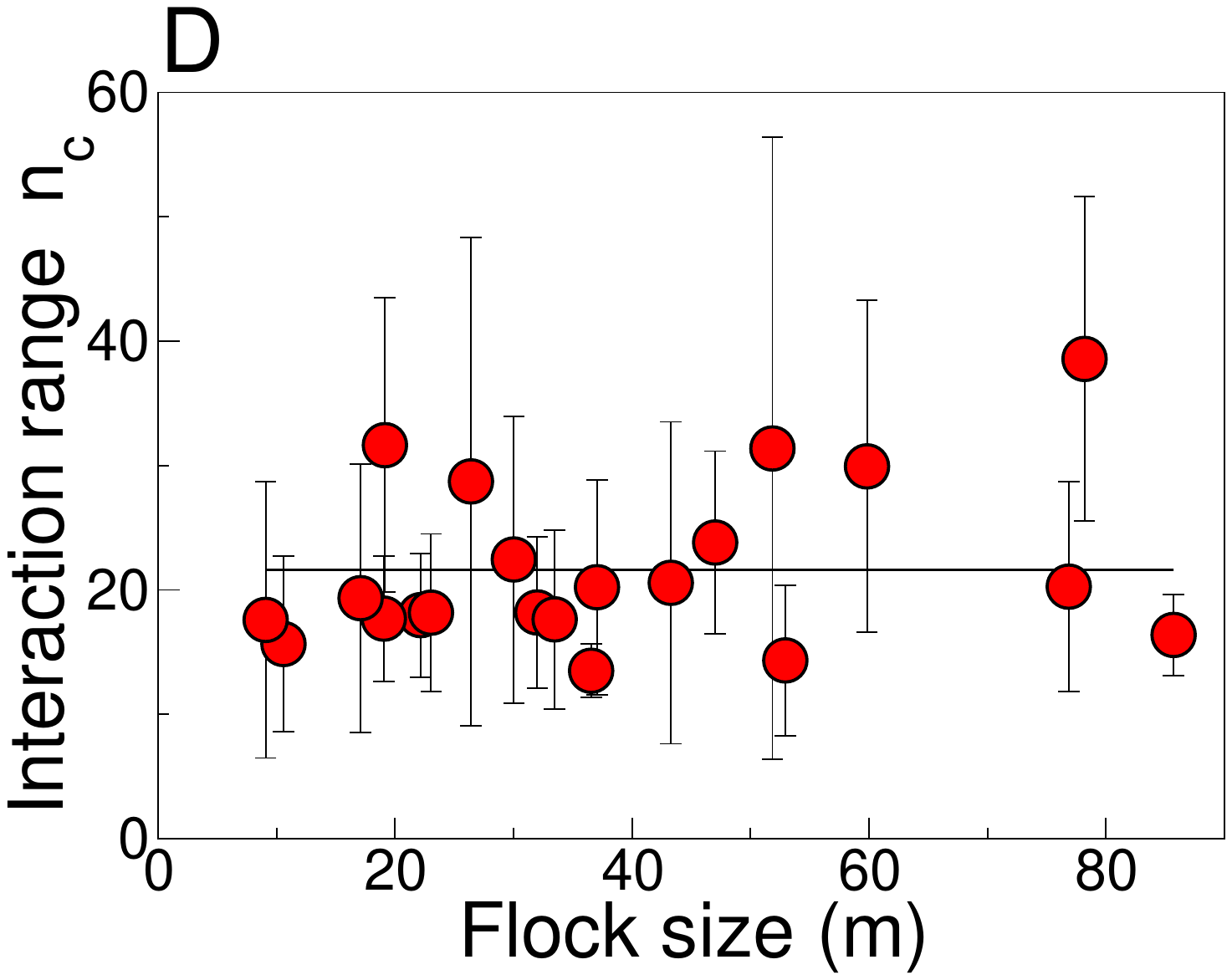}
\includegraphics[width = 0.3 \linewidth]{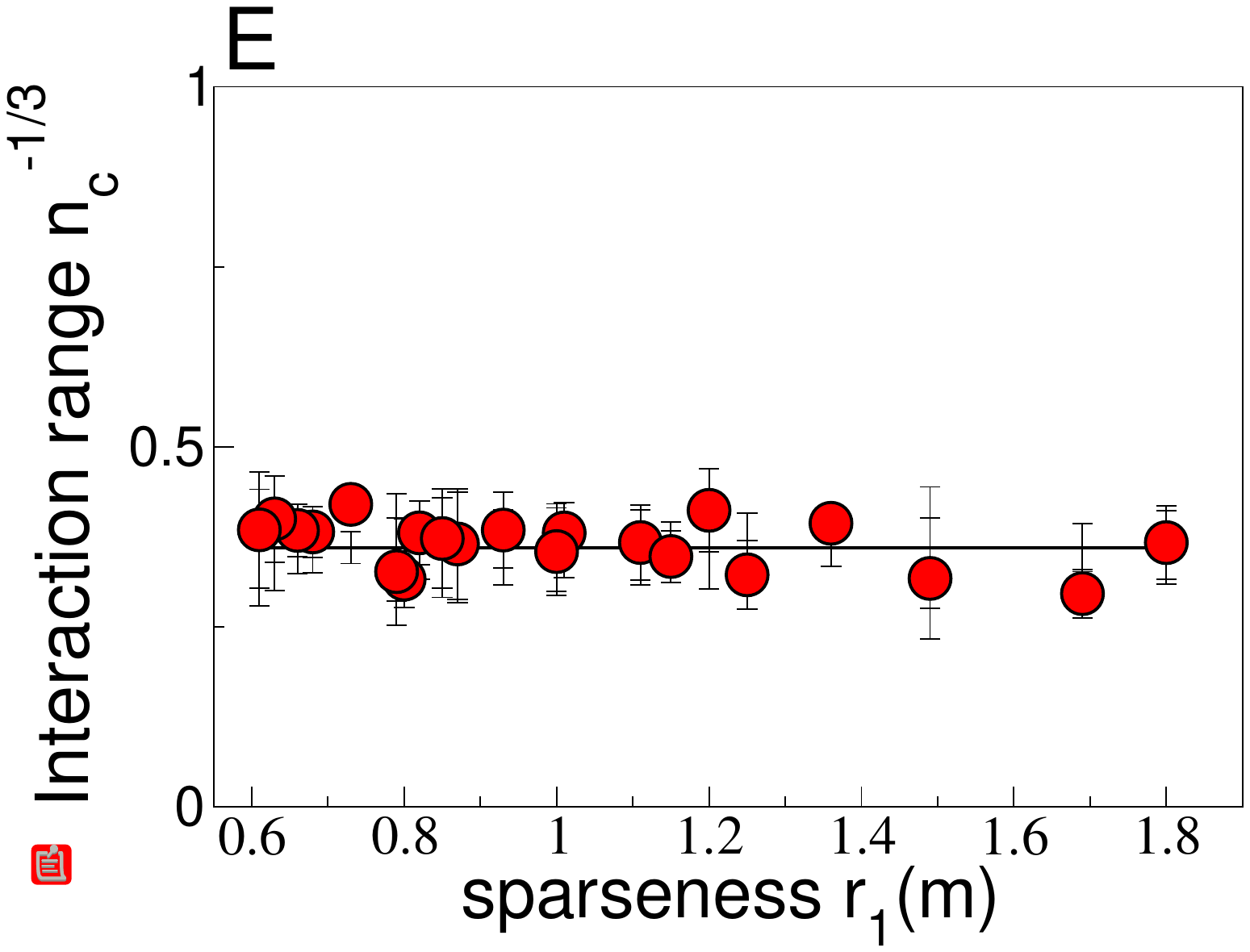} 
\includegraphics[width = 0.3 \linewidth]{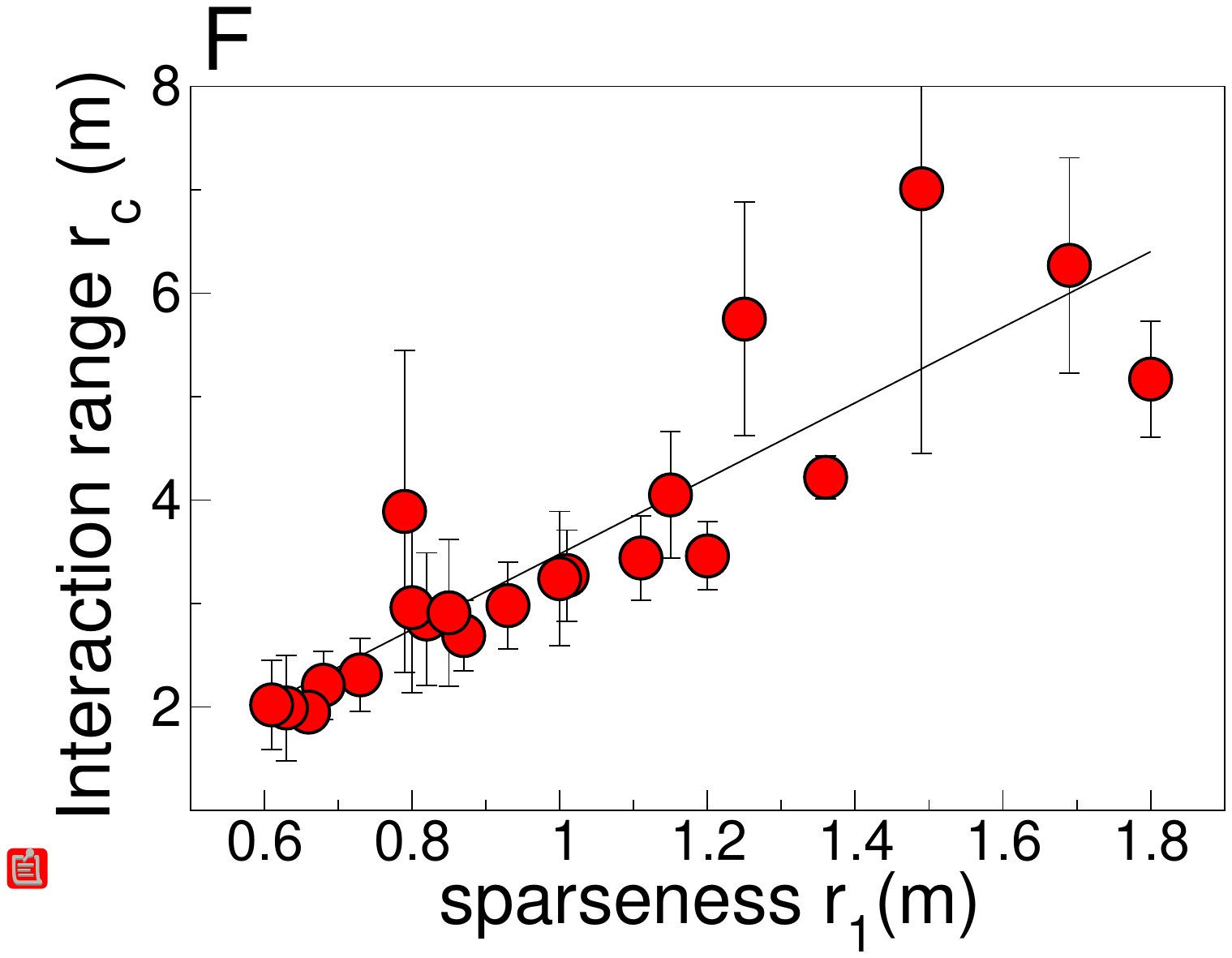}
\caption{Setting the strength and range of interactions. 
(A) The predicted strength of correlation, $C_{\rm int}$, as a function of the interaction strength $J$, with $n_c = 11$, for the snapshot  in Fig~\ref{vectors}.   Matching the experimental value of $C_{\rm int}= 0.99592$ determines $J = 45.73$. Inset: Zoom of the crossing point; error bars are obtained from the model's predictions of fluctuations of $C_{\rm int}(J,n_c)$.
(B) The log--likelihood of the data per bird ($\langle \ln P(\{{\vec s}_{\rm i}\})\rangle_{\rm exp}/N$) as a function of $n_c$ with $J$ optimized for each value of $n_c$; same snapshot as in (A).  There is a clear maximum at $n_c = 11$. Inset: the log--likelihood per bird for other snapshots of the same flocking event. 
(C)  The inferred value of $J$ for all observed flocks, shown as a function of the flock's size. Each point corresponds to an average over all the snapshots of the same flock. Error bars are standard deviations across multiple snapshots. 
(D) As in (C), but for the inferred values of $n_c$. Averaging  over all flocks we find $n_c=21.2\pm 1.7$ (black line). 
(E) The inferred value of the topological range $n_c^{-1/3}$ as a function of the mean inter--bird distance in the flock, for all flocks.   Error bars are standard deviations across multiple snapshots of the same flock.  (F) As in (E), but for the metric range $r_c$.  Interactions extend over  some fixed metric distance $r_0$, then we expect $n_c^{-1/3} \propto r_1/r_0$ and $r_c = {\rm constant}$; we find the opposite pattern, which is a signature of interactions with a fixed number of topological neighbors \cite{ballerini+al_08b}.\label{fit}}
\end{figure*}

Importantly, with large flocks we can estimate the correlations among interacting neighbors from a single snapshot of the birds' flight directions $\{{{\vec s}_{\rm i}}\}$, as indicated in the second step of Eq (\ref{corre_local}).  In contrast, if we were trying to estimate the entire correlation matrix in Eq (\ref{unzu}), we would need as many samples as we have birds in the flock (see Appendix \ref{App:spinwave}), and we would have to treat explicitly the dynamic rearrangements of the interaction network during flight.    This is an extreme version of the general observation that the sampling problems involved in the construction of maximum entropy models can be greatly reduced if we have prior expectations that constrain the structure of the interaction matrix \cite{cocco+monasson_11,cocco+al_11}.

We now apply this analysis to data on real flocks of starlings.  Given a snapshot of the flock, we have the configuration $\{{\vec s}_{\rm i}\}$, and we need to evaluate the probability  $P(\{{\vec s}_{\rm i}\})$ in Eq (\ref{maxent_local}) for any value of $J$ and $n_c$,  then maximize this probability with respect to these parameters (see Methods and the Appendices for details of the calculation). Special care must be devoted to birds on the outer edge, or border, of the flock, since these individuals have very asymmetric neighborhoods and may receive inputs from signals outside the flock.  If we take the flight directions of these border birds as given, we can study how information propagates through the flock, without having to make assumptions about how the boundary is different from the interior.  Technically, then, we describe the flock with Eq (\ref{maxent_local}), but with the flight directions of the border birds fixed (again, see Methods and especially Appendix \ref{App:freebound} for details).

We proceed as follows:  For a single flock, at a given instant of time, we compute the correlation $C_{\rm int}$ predicted by the model in Eq (\ref{maxent_local}) as a function of the coupling strength $J$,  and compare it with the experimental value of the correlation (Fig \ref{fit}A). The equation $C_{\rm int}(J,n_c) = C^{\rm exp}_{\rm int}$, fixes $J(n_c)$ for each value of $n_c$. Then we fix the interaction range by looking at the overall probability of the data as a function of $n_c$. In general there is a clear optimum (Fig \ref{fit}B), from which we finally infer the maximum entropy value of both parameters, $n_c$ and $J$.  We repeat this procedure for every snapshot of each flock, and compute the mean and standard deviation of the interaction parameters for each flock over time.   Alternatively,  we can average  the log--likelihood over many snapshots, and then optimize, and this gives equivalent results for $J$ and $n_c$ (see Appendix \ref{App:global}). 

In Figures \ref{fit}C and D we report the value of the interaction strength $J$ and of the interaction range $n_c$ for all flocks, as a function of the flock's  spatial size, $L$. The inferred values of $J$ and $n_c$ are reproducible, although error bars are larger for smaller flocks.  In particular, $J$ and $n_c$ do not show any significant trend with the flocks' linear dimensions, with the number of birds, or with the density.   This did not have to be the case, nor is it in any way built in to our framework; for example, if the real interactions extended over long distances, then our fitting procedure would produce an increase of $n_c$ and $J$ with the size of the flock.

\begin{figure*}
\includegraphics[width = 0.4\linewidth]{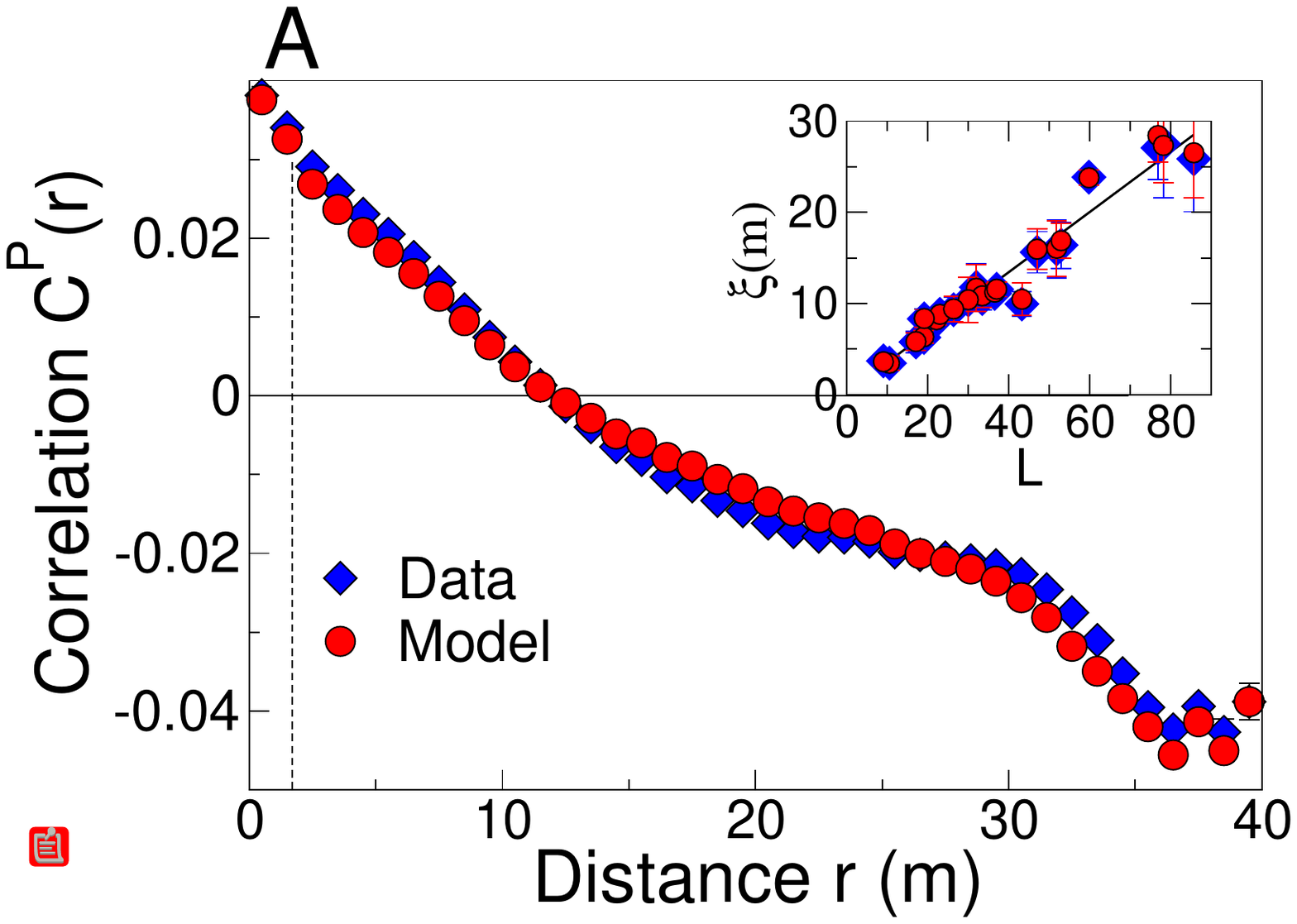}
\includegraphics[width = 0.4\linewidth]{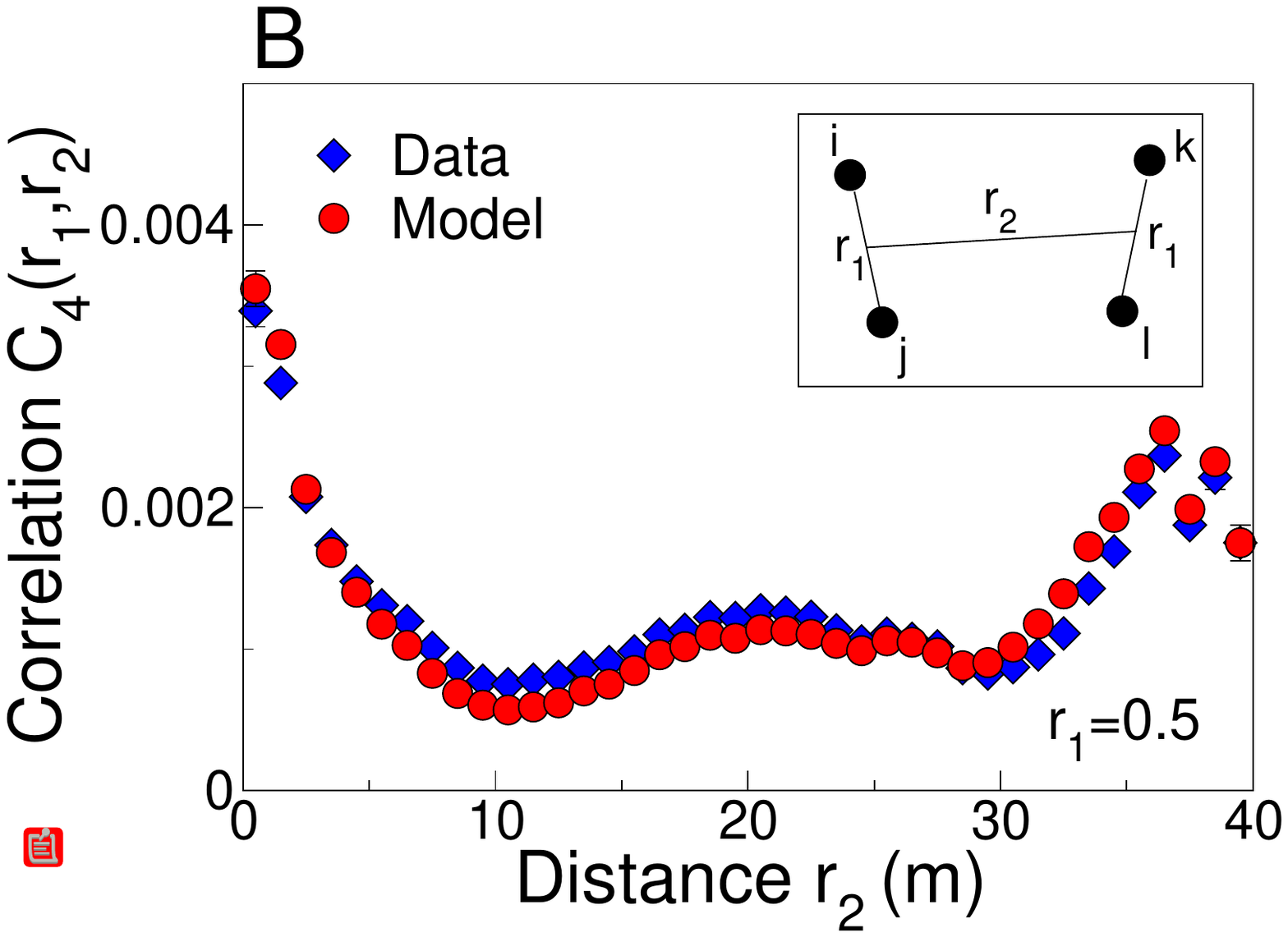}
\includegraphics[width = 0.4\linewidth]{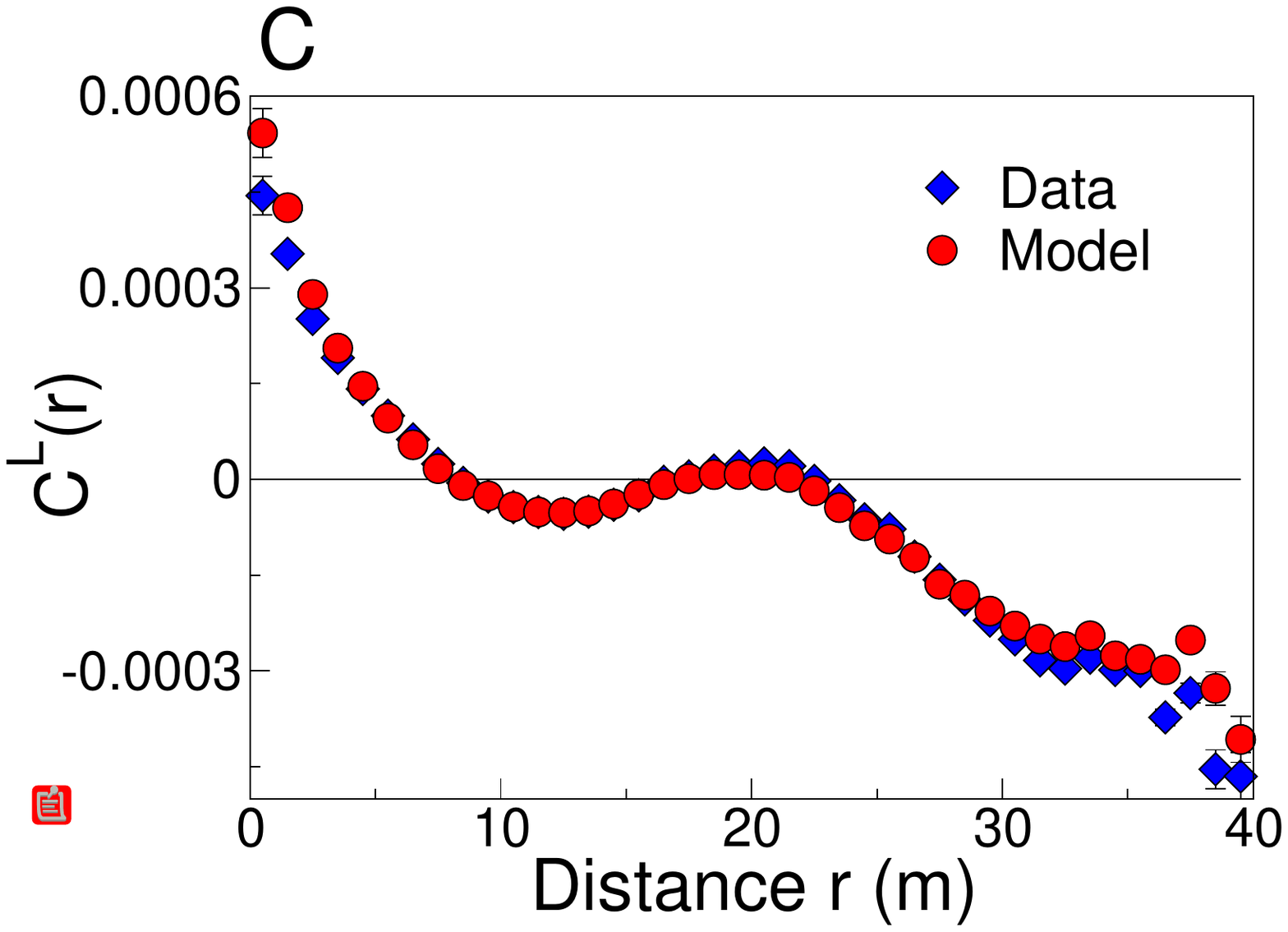}
\includegraphics[width = 0.4\linewidth]{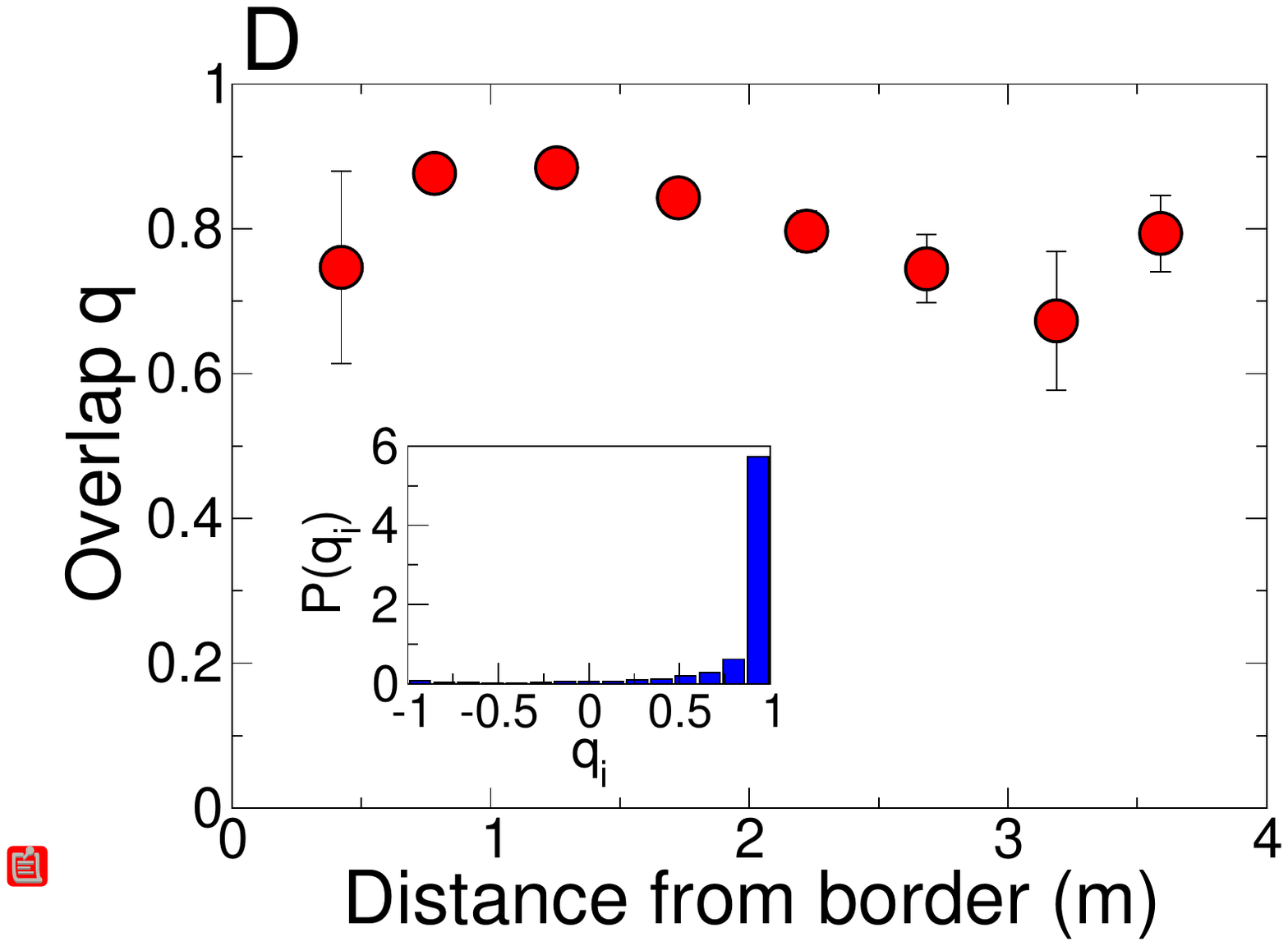}
\caption{Correlation functions predicted by the maximum entropy model vs experiment. The full pair correlation function can be written in terms of a longitudinal and a perpendicular component, i.e. $\langle {\bf {\vec s}}_{\rm i}\cdot {\bf {\vec s}}_{\rm j} \rangle = \langle s_{\rm i}^{\rm L}s_{\rm j}^{\rm L} \rangle + \langle {\vec \pi}_{\rm i} \cdot {\vec \pi}_{\rm j}\rangle$. Since the two components have different amplitudes, it is convenient to look at them separately.
(A) Perpendicular component of the correlation, $C^{\rm P}(r)=\langle {\vec \pi}_{\rm i} \cdot {\vec\pi}_{\rm j}\rangle$, as a function of the distance;  the average is performed over all pairs $\rm ij$ separated by distance $r$.  Blue diamonds refer to experimental data (for the sample in Fig \ref{vectors}), red circles to the prediction of the model in Eq (\ref{maxent_local}). The dashed line marks the maximum $r$ that contributes to $C_{\rm int}$, which is the only input to the model. The correlation function is well fitted over all length scales. In particular, the correlation length $\xi$, defined as the distance where the correlation crosses zero, is well reproduced by the model. Inset: $\xi$ vs. size of the flock, for all the  flocking events; error bars are standard deviations across multiple snapshots of the same flocking event. 
(B)  Four--point correlation function $C_{4}(r_1;r_2)=\langle \left ({\bf\vec \pi}_{\rm i}{\bf \cdot} {\bf\vec \pi}_{\rm j} \right )\left ({\bf\vec \pi}_{\rm k}{\bf \cdot} {\bf\vec \pi}_{\rm l} \right )\rangle$, where  the pairs $\rm ij$ and $\rm kl$ are as shown in the inset  (see also Appendix \ref{App:correlations}). The figure shows the behaviour of $C_{4}(r_1;r_2)$ as a function of $r_2$, with $r_1=0.5$.  
(C) Longitudinal component of the correlation $C^{\rm L}(r)=\langle s_{\rm i}^{\rm L} s_{\rm j}^{\rm L}\rangle-S^2$, as a function of distance.   Note that in the spin wave approximation, $C^{\rm L}(r)=1-C_{4}(0;r)-S^2$. 
(D) Similarity between the predicted mean value of flight direction, $\langle \vec{\pi}_{\rm i}\rangle$, and real data, for all individual birds in the interior of the flock.  The similarity can be quantified through the local overlap $q_i= \langle {\vec\pi}_{\rm i}\rangle \cdot {\vec \pi}^{\rm exp}_{\rm i}/(|\langle {\vec\pi}_{\rm i}\rangle ||{\vec\pi}^{\rm exp}_{\rm i}|)$, which is plotted as a function of the distance of the individual from the border. Maximal similarity corresponds to $q_{\rm i}=1$. Inset: full distribution $P(q)$ for all the interior birds.
\label{correlations}}  
\end{figure*}

In Figure~\ref{fit}E we also show that the interaction range $n_c$ does not depend on the typical distance between neighboring birds, $r_1$, which is closely related to the flock's density. Of course, we can run exactly the same method using a {\em metric} interaction range, $r_c$, rather than a {\em topological} range, $n_c$. We simply set $J_{\rm ij}=J$ if and only if birds $\rm  i$ and $\rm j$ lie within $r_c$ meters. When we do this we find that the metric range $r_c$ does depend on the nearest neighbor distance $r_1$, in contrast with the topological range $n_c$ (Fig \ref{fit}F). This result provides strong support for the claim put forward in Ref \cite{ballerini+al_08b} that birds interact with a fixed number of neighbors, rather than with all the birds within a fixed metric distance.

Having fixed $J$ and $n_c$ by matching the scalar correlation in the flock, we have no free parameters---everything that we calculate now is a parameter free prediction.  We begin by computing the correlations between pairs of birds as a function of their distance, $C(r)\sim \sum_{\rm ij} {{\vec s}_{\rm i}} {\bf \cdot} {{\vec s}_{\rm j}}\; \delta(r-r_{\rm ij})$, as shown in  Fig \ref{correlations}A.  There is extremely good agreement across the full range of distances. As we have seen,  our maximum entropy calculation finds local interactions, i.e. a relatively small value of $n_c$ ($n_c \sim 20$ for flocks of up to thousands birds). This implies that the scalar correlation $C_{\rm int}$, used as an experimental input to the calculation, is the integral of $C(r)$ only over a very small interval close to $r=0$: only the average value of pair correlations at very short distances is used as an input to the calculation, whereas all the long range part of $C(r)$ is not. Nevertheless, we have very good agreement out to the overall extent of the flock itself. This confirms our expectation that a model for the local correlations is able to describe correlations on all length scales.   We draw attention to the fact that the apparent correlation length, defined by $C (r=\xi) = 0$, is predicted to scale with the linear size of the flock ($\xi\propto L$, inset to Fig \ref{correlations}A), as observed experimentally.

Correlations exist not just between pairs of birds, but among larger $n$--tuplets.  In Fig \ref{correlations}B we consider the correlations among quadruplets of birds. Although these correlations are small, their shape is nontrivial and quite noiseless. The model, which takes only local pairwise correlations as input, reproduces very accurately these $4$-body correlations, including a non--monotonic dependence on distance, out to distances comparable to the full extent of the flock.  Again, this is not a fit, but a parameter free prediction.   

Finally, instead of measuring correlations, we can ask the model to predict the actual flight  directions of individual birds in the interior of the flock, given the directions chosen by birds on the borders.  This can't work perfectly, since the model predicts that individual birds have an element of randomness in their choice of direction relative to their neighbors, but as shown in Fig \ref{correlations}D the overlap between predicted and observed directions is very good,  not just for birds close to the border but throughout the entire ``thickness'' of the flock.

\begin{figure}
\includegraphics[width = 0.49\linewidth]{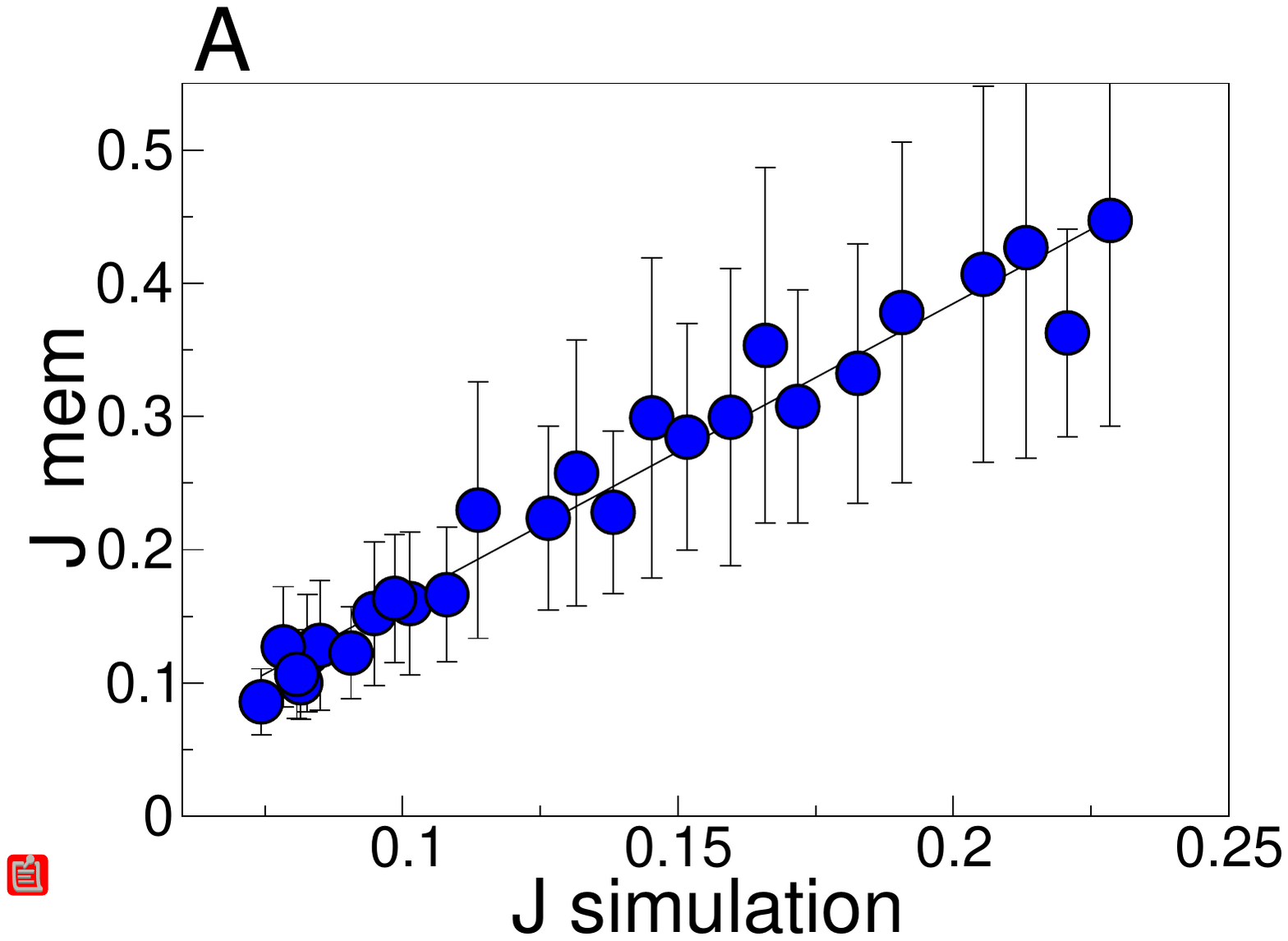}
\includegraphics[width = 0.49\linewidth]{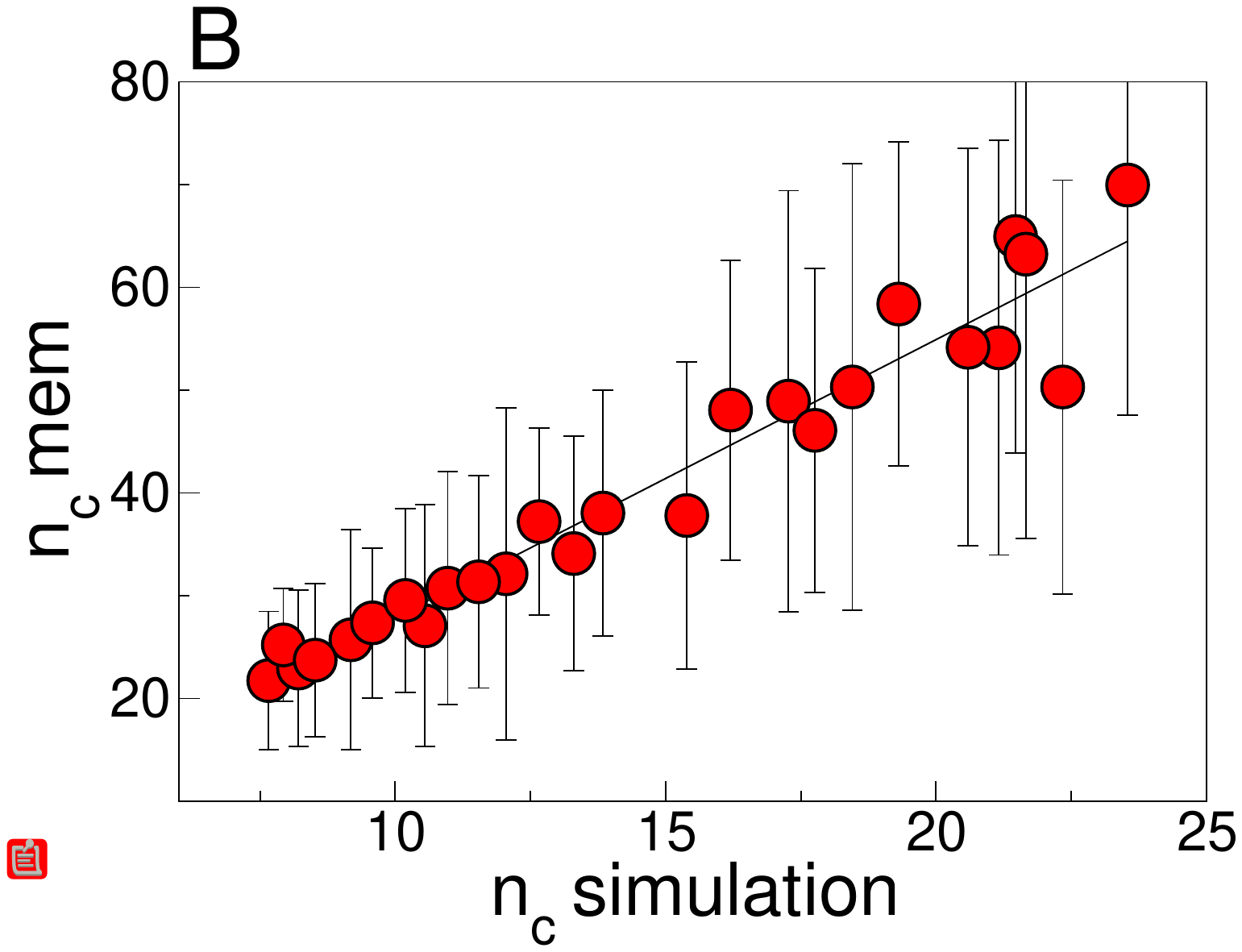}
\caption{Maximum entropy analysis for a model of self--propelled particles. 
(A) Inferred value of the parameter $J$ vs. microscopic strength of alignment forces used in the simulation. 
(B) Inferred value of $n_c$ vs. the true number of interacting neighbors in the simualtion.  Slopes of the lines are 2.2 and 2.7, respectively. Error bars are standard deviations across 45 snapshots of the same simulation. 
\label{fig_model}}
\end{figure}

The maximum entropy model has a mechanistic interpretation, from Eq (\ref{langevin}), in terms of social forces driving the alignment of the flight directions.  Given the success of the model in predicting the propagation of order throughout the flock, it is interesting to ask whether we can take this mechanistic interpretation seriously.  As a test, we have simulated a population of self--propelled particles in three dimensions moving according to social forces that tend to align each particle with the average direction of its neighbors, as described by Eqs. (\ref{model_vel}) and (\ref{model_pos}) in the Methods.  We then compared the simulation parameters ($J^{\rm sim}, n^{\rm sim}_c)$ to the values $(J^{\rm mem}, n^{\rm mem}_c)$ obtained by applying the maximum entropy method to snapshots drawn from the simulation, just as we have analyzed the real data.  Both the strength and the range of the interaction given by the maximum entropy analysis are proportional to the ``microscopic" parameters used in the simulation (Figs ~\ref{fig_model}A and B), although the maximum entropy interaction range $n^{\rm mem}_c$ is roughly $3\times$ larger than the true number of interacting neighbors, $n_c^{\rm sim}$. We believe that this overestimation is due to the fact that birds (unlike spins) move through the flock, encountering new neighbors before losing memory of the earlier flight directions, and in so doing propagate information and correlation more effectively than if they were sitting on a fixed network.  In other words, the maximum entropy model, where interactions are static by construction, compensates the dynamical nature of the true interaction network by giving a larger effective value of $n_c$.  Hydrodynamic theories of flocking \cite{toner+tu_95,toner+tu_98} provide an analytic treatment of this effect, which is essential for collective motion of large two--dimensional groups. Indeed, in the limit of very large flocks, this ratio between the microscopic range of interactions and the effective range recovered by maximum entropy methods could become arbitrarily large,  but the flocks we study here seem not to be big enough for this effect to take over.    If we use the ``calibration'' of model from Fig \ref{fig_model}B, then the observation of $n_c^{\rm } =21.6$ in the real flocks (Fig \ref{fit}) suggests that the true interactions extend over $n_c= 7.8$,  in reasonable agreement with the result from \cite{ballerini+al_08b,tavarone_09}, $n_c=7.0\pm 0.6$ using very different methods.

To summarize, we have constructed the minimal model that is consistent with a single number characterizing the interactions among birds in a flock, the average correlation between the flight directions of immediate neighbors.  Perhaps surprisingly, this provides  an essentially complete theory for the propagation of directional order throughout the flock, with no free parameters.  The theory predicts major qualitative effects, such as the presence of long ranged, scale free correlations among pairs of birds, as well as smaller, detailed effects such as the non--monotonic distance dependence of (four--point) correlations among two pairs of birds.  The structure of the model corresponds to interactions with a fixed number of (topological) neighbors, rather than with all neighbors that fall within a certain (metric) distance; the relevant number of neighbors and the strength of the interaction are remarkably constant across multiple flocking events.   Our approach can be seen as part of a larger effort   using maximum entropy methods to link the collective behaviour of real biological systems to theories grounded in statistical mechanics \cite{schneidman+al_06,tkacik+al_06,shlens+al_06,psu_group,tkacik_07,bialek+ranganathan_07,tang+al_08,yu+al_08,tkacik+al_09,weigt+al_09,halabi+al_09,mora+al_10,stephens+bialek_10}.  As with these other examples, we view the success of our theory as an encouraging first step.  We have focused on the flight directions, taking the positions of the birds as given.  A full theory must connect the velocities of the birds to their evolving positions, which requires more accurate measurements of trajectories over time, and we must consider the fluctuations in the speed as well as direction of flight.  There are maximum entropy approaches to both of these problems, and the mapping from maximum entropy models to statistical mechanics suggests that the observation of scale free correlations in speed fluctuations \cite{cavagna+al_09} will locate models of flocking at an especially interesting point in their parameter space \cite{mora+bialek_11}.

{\bf Acknowledgements.}  We thank I Couzin, T Grigera, N Leonard, G Parisi, G Theraulaz and J Toner for helpful discussions.  Work in Princeton was supported in part by NSF Grants IIS--0613435 and PHY--0957573; work in Rome was supported in part  by Grants 
IIT--Seed Artswarm, ERC--StG n. 257126 and AFOSR--Z809101. Our collaboration was facilitated by the Initiative for the Theoretical Sciences at the Graduate Center of the City University of New York, supported in part by the Burroughs--Wellcome Fund.

\section*{Methods}

{\sl \bf Data.}  Analyzed data were obtained from experiments on large flocks of starlings ({\em Sturnus vulgaris}), in the field. Using stereometric photography and innovative computer vision techniques \cite{cavagna+al_08a,cavagna+al_08b} the individual 3D coordinates  and velocities  were measured in cohesive groups of up to 4268 individuals \cite{ballerini+al_08b,cavagna+al_09,ballerini+al_08a}. The data-set comprises 21 distinct flocking events (see Table  \ref{table}). Each event consists of up to 40   consecutive 3D configurations (individual positions and velocities), at time intervals of 1/10 s.

{\sl \bf Analytic approach to the maximum entropy model.}  To apply the maximum entropy analysis, we need to compute the expected values of correlation functions according  to the measure defined in Eq (\ref{maxent1}). This requires the computation of the partition function $Z(\{J_{\rm ij}\})$, which is, in general, a hard task. Flocks, however, are very ordered groups, in that birds' velocities are neatly aligned to each other \cite{cavagna+al_09}. In this case we can use the  ``spin wave'' approximation \cite{ref-spinwave}, which exploits the strong ordering condition.  Let us call ${\vec S}=(1/N) \sum_{\rm i}{\vec s}_{\rm i}=S \  {\vec n}$ the global order parameter, or polarization, measuring the degree of collective alignment, where $\vec{n}$ is a unit vector. Individual orientations can be rewritten in terms of a {\it longitudinal} and a {\it perpendicular} component: ${\vec s}_{\rm i} = s_{\rm i}^{\rm L} {\vec n} + {\vec{\bf\pi}}_{\rm i}$.  
If the system is highly polarized, $S\sim 1$,  $|{\vec \pi}_{\rm i}| \ll 1$, and $s_{\rm i}^L\sim1- |\vec{\pi}_{\rm i}|^2/2$; we verify this last condition for our data in  Fig \ref{check_spinwave} of Appendix \ref{App:spinwave}.  The partition function can be written as an integral over the  $\{{\vec \pi}\}$, and if $S\sim 1$ the leading terms are (see Appendix \ref{App:spinwave} for details):
\begin{equation}
Z(\{J_{\rm ij}\}) \approx \int d^N \vec{\mbox{\boldmath$\pi$} } \exp{\left [ -\frac{1}{2} \sum_{\rm i,j =1}^N A_{\rm ij} {\vec \pi}_{\rm i}\cdot{\vec\pi}_{\rm j}+\frac{1}{2}\sum_{\rm i,j=1}^{N}J_{\rm ij}\right ]}   ,
\label{spinwave_met}
\end{equation}
where $A_{\rm ij}=\sum_k J_{\rm ik} \delta_{\rm ij} - J_{\rm ij}$, $ d^N \vec{\mbox{\boldmath$\pi$} }=\prod_{\rm i} d\vec{\pi}_{\rm i}$ and the $\{{\vec\pi}_{\rm i}\}$ satisfy the constraint $\sum_i {\vec \pi}_{\rm i}=0$ . If we consider the flight directions of birds on the border as given, integration must be performed with respect to internal variables only (see Appendix \ref{App:fixedbound}). 
After some algebra  one gets 
\begin{widetext}
\begin{equation}
Z(\{J_{\rm ij}\};{\cal B})= \int d^{\cal I} \vec{\mbox{\boldmath$\pi$} }
\exp \left [ -\frac{1}{2} \sum_{{\rm i,j}\in{\cal I}} A_{\rm ij} {\vec \pi}_{\rm i}\cdot{\vec\pi}_{\rm j}+\sum_{{\rm i} \in {\cal I}} {\vec\pi}_{\rm i}\cdot  {\vec h}_{\rm i}  \nonumber + \frac{1}{2}\sum_{{\rm i,j}\in{\cal I}} J_{\rm ij} +\frac{1}{2}\sum_{{\rm i} \in{\cal I}}{h}^{\rm L}_{\rm i}+\frac{1}{2}\sum_{{\rm i,j}\in{\cal B}}J_{\rm i,j}{\vec s}_{\rm i}\cdot{\vec s}_{\rm j} \right ] \ .
\label{z_border}
\end{equation}
\end{widetext}
Here ${\cal I}$ and ${\cal B}$ represent the subsets of, respectively, internal and border individuals; ${\vec h}_{\rm i} = \sum_{{\rm l}\in {\cal B}} J_{\rm il} {\vec s}_{\rm l}$ is a `field' describing the influence of birds on the border on internal bird i; and, now, $A_{\rm ij} = \delta_{\rm ij} (\sum_{k\in{\cal I}} J_{\rm ik}+\sum_{l\in{\cal B} }J_{\rm il} {s}^{\rm L}_{\rm l})$.  The integral in Eq (\ref{z_border}) can be carried out explicitly; see Eq (\ref{Z_border2}). The  reduced model in Eq (\ref{maxent_local}) corresponds to  $J_{\rm ij}= J\ n_{\rm ij}$, with $n_{\rm ij} =1$, $1/2$, or $0$ according to whether both individuals, just one, or none, belong to the local $n_c$-neighborhood of the other.   Given the individual coordinates of birds in space, the matrix $A_{\rm ij}$ can  be computed for any given snapshot, and  $Z(J,n_c;{\cal B})$ (and correlation functions) can be calculated as a function of $J$ and $n_c$.  These two parameters must then be adjusted to maximize the log--likelihood of the data,
\begin{equation}
{\bigg\langle}\log P\left(\{{\vec s}_{\rm i}\}\right){\bigg\rangle}_{\rm exp}=-\log Z(J,n_c;{\cal B}) +\frac{1}{2} J n_c N C^{\rm exp}_{\rm int} \ .
\label{loglikelihood}
\end{equation}
Maximizing with respect to $J$ corresponds to equating expected and experimental correlations. In our case, this equation can be solved analytically, leading to an explicit expression of  the optimal $J$ vs. $n_c$; see Eq (\ref{final_J}).  Minimization with respect to $n_c$ can then be performed numerically.  A graphical visualization of the solution can be found in Fig.~\ref{fit}.

{\sl \bf Self-propelled particle model.}   We consider a model of self--propelled particles extensively studied in the literature \cite{gregoire+chate_04}. Each particle moves with vector velocity $\vec v_{\rm i}(t)$ according to the following equations:
\begin{eqnarray}
&&{\vec v}_{\rm i}(t+1)=v_0\Theta \left [ \alpha \sum_{ {\rm j} \in n_c^{\rm i} } {\vec v}_{\rm j} (t) + \beta \sum_{ {\rm j} \in n_c^{\rm i} } {\vec f}_{\rm ij} + n_c \ {\vec \eta}_{\rm i}\right ] \label{model_vel} \\
&&{\vec x}_{\rm i}(t+1)={\vec x}_{\rm i}(t)+{\vec v}_{\rm i}(t)
\ ,
\label{model_pos}
\end{eqnarray}
where $\Theta$ is a normalization operator $\Theta({\vec y})={\vec y}/|{\vec y}|$ that serves to keep the speed fixed at $|{\vec v} | = v_0$, and ${\rm j} \in {n_c^{\rm i}}$ means that $\rm j$ belongs to the $n_c$ interacting  neighbors of $\rm i$.  The distance--dependent force ${\vec f}_{\rm ij}$   acts along the direction  connecting $\rm i$ and $\rm j$; following Ref \cite{gregoire+chate_04}, if $\vec{e}_{\rm ij}$ is the unit vector between i and j, we take
\begin{eqnarray}
\vec{f}_{\rm ij} (r_{\rm ij} < r_{b}) &=& - \infty \ \vec{e}_{\rm ij}\\
\vec{f}_{\rm ij} (r_b<r_{\rm ij}<r_a)  &=& {1\over 4}\cdot {{  r_{\rm ij} - r_e}\over {r_a-r_e}}\vec{e}_{\rm ij}\\
\vec{f}_{\rm ij} (r_a<r_{\rm ij}<r_0) &=& \ \vec{e}_{\rm ij} ,
\end{eqnarray}
Finally, ${\vec \eta}_{\rm i}$ is a random unit vector, independent for each bird and at each moment of time.  The parameters $\alpha$ and $\beta$ tune the strength of the alignment and of the cohesion force, respectively; in particular, the strength of alignment is given by $J=v_0 \alpha/n_{c}$.   To test the maximum entropy analysis, we modified the model in such a way that we could vary $n_c$. Specifically, we introduced an angular resolution $\mu$ such that only neighbors with mutual angles larger than $\mu$ were included in the neighborhood. When $\mu$ is of the order of the Voronoi angle the model is statistically equivalent to the original version (where Voronoi neighbors were considered), but increasing (decreasing) $\mu$ one can decrease (increase) the value of $n_c$. In this way  both the number $n_c$ of interacting neighbors and the strength of the interaction $J$ can be arbitrarily tuned.  Parameters  were chosen as $r_0 = 1$ (to set the scale of distance), $r_b = 0.2$, $r_e = 0.5$, $r_a = 0.8$, $\alpha=35$, $\beta=5$, $v_0=0.05$, and we simulated a flock of $N=512$ birds.

\begin{table}[h]
\vskip 0.1 in
\begin{tabular}{|c|c|c|c|c|}
\hline
 Event & $N$ & $S$ & $v_0$ (${\rm m}/{\rm s}$) & $L$ ($\rm m$) \\
 \hline
 17--06 & 552 & 0.935 & 9.4 & 51.8\\
 \hline
 21--06 & 717 & 0.973 & 11.8  & 32.1\\
 \hline
 25--08 & 1571 & 0.962 & 12.1 & 59.8\\
 \hline
 25--10 & 1047 & 0.991 & 12.5 & 33.5\\
 \hline
 25--11 & 1176 & 0.959 & 10.2 & 43.3\\
 \hline
 28--10 & 1246 & 0.982 & 11.1 & 36.5\\
 \hline
 29--03 & 440 & 0.963 & 10.4 & 37.1\\
\hline
31--01 & 2126 & 0.844 & 6.8 & 76.8\\
\hline
32--06 & 809 & 0.981 & 9.8 & 22.2\\
\hline
42--03 & 431 & 0.979 & 10.4 & 29.9\\
\hline
49--05 & 797 & 0.995 & 13.9 & 19.2\\
\hline
54--08 & 4268 & 0.966 & 19.1 & 78.7\\
\hline
57--03 & 3242 & 0.978 & 14.1 & 85.7\\
\hline
58--06 & 442 & 0.984 & 10.1 & 23.1\\
\hline
58--07 & 554 & 0.977 & 10.5 & 19.1\\
\hline
63--05 & 890 & 0.978 & 9.9 & 52.9\\
\hline
69--09 & 239 & 0.985 & 11.8 & 17.1\\
\hline
69--10 & 1129 & 0.987 & 11.9 & 47.3\\
\hline
69--19 & 803 & 0.975 & 13.8 & 26.4\\
\hline
72--02 & 122 & 0.992 & 13.2 & 10.6\\
\hline
77--07 & 186 & 0.978 & 9.3 & 9.1\\
\hline 
\end{tabular}
\caption{Summary of experimental data. Flocking events are labelled according to experimental session number and to the position within the session they belong to. The number of birds $N$ is the number of individuals for which we obtained a 3D reconstruction of positions in space. The polarization $S$  is defined in the Methods  and in Appendix \ref{App:spinwave}. The linear size $L$ of the flock is defined as the maximum distance between two birds belonging to the flock. The speed $v_0$ is that  of the centre of mass, i.e. the mean velocity of the group. All values are averaged over several snapshots during the flocking event. \label{table}}
\end{table}

\appendix

\section{Maximum entropy approach}
\label{App:maxent}

The maximum entropy method has a long history.  Recent developments in experimental methods have renewed interest in this idea as a path for constructing statistical mechanics models of biological systems directly from real data, with examples drawn from networks of neurons \cite{schneidman+al_06,tkacik+al_06,shlens+al_06,tang+al_08,yu+al_08,tkacik+al_09}, ensembles of amino acid sequences \cite{bialek+ranganathan_07,weigt+al_09,halabi+al_09,mora+al_10}, biochemical and genetic networks \cite{psu_group,tkacik_07}, and the statistics of letters in words \cite{stephens+bialek_10}.  Here we give a review of the basic ideas leading to Eq (\ref{maxent1}) of the main text, hoping to make the discussion accessible to a wider readership.

Imagine a system whose state at any one instant of time is described by a set of variables $\{x_1 ,\, x_2 ,\, \cdots , \, x_N\} \equiv \bf x$.  For the moment we don't need to specify the nature of these variables---they could be positions or velocities of individual birds ${\rm i} = 1, 2, \cdots N$ in a flock, or more subtle parameters of body shape or instantaneous posture.  Whatever our choice of variables, we know that when the number of elements in the system $N$ (here, the number of birds in the flock)  becomes large, the space $\bf x$ becomes exponentially larger.  Thus there is no sense in which we can ``measure'' the distribution of states taken on by the system, because the number of possibilities is just too large.  On the other hand, we can obtain reliable measurements of certain average quantities that are related to the state $\bf x$.  To give a familiar example, we can't measure the velocity of every electron in a piece of wire, but certainly we can measure the average current that flows through the wire.  Formally, there can be several such functions,  $f_1({\bf x}), \, f_2({\bf x}),\, \cdots ,\, f_K({\bf x})$, of the state ${\bf x}$. The minimally structured distribution for these data is the most random distribution $P({\bf x})$ that is consistent with the observed averages of  these functions $\{\langle f_\nu({\bf x})\rangle_{\rm exp}\}$, where $\langle \cdots \rangle_{\rm exp}$ denotes an average measured experimentally.

To find the ``most random'' distribution, we need a measure of randomness.  Another way to say this is that we want the distribution $P({\bf x})$ to hide as much information about ${\bf x}$ as possible.  One might worry that information and randomness are qualitative concepts, so that there would be many ways to implement this idea.  In fact, Shannon proved that there is only one measure of randomness or available information that is consistent with certain simple criteria \cite{shannon_48,cover+thomas_91}, and this is the entropy
\begin{equation}
S\left [P\right ]=-\sum_{\bf x} P({\bf x})\ln P({\bf x}) \ .
\label{entropy}
\end{equation}
Thus we want to maximize $S\left [P\right ]$ subject to the constraint that the expectation values computed with $P$ match the experimentally measured ones, that is  
\begin{equation}
\langle f_\mu ({\bf x})\rangle_{\rm exp} = \langle f_\mu ({\bf x})\rangle_P \equiv \sum_{\bf x} P({\bf x}) f_\mu ({\bf x}) 
\end{equation}
for all $\mu$ \cite{jaynes_57}.  The distribution $P({\bf x})$ must also be normalized, and it is convenient to think of this as the statement that the average of the ``function'' $f_0 ({\bf x}) = 1$ must equal the ``experimental'' value of $1$.  Our constrained optimization problem can be solved using the method of the Lagrange multipliers \cite{bender+orszag}:  we introduce a generalized entropy function,
\begin{equation}
{\cal S}\left[ P;\{\lambda_\nu\}\right]=S\left [P\right ]-\sum_{\mu =0}^K  \lambda_{\mu} \left [ \langle f_{\mu}({\bf x})\rangle_P - \langle f_{\mu}({\bf x})\rangle_{\rm exp} \right ] \ ,
\label{gen_entropy}
\end{equation}
where a multiplier $\lambda_\mu$ appears for each constraint to be satisfied, and then we maximize $\cal S$ with respect to the probability distribution $P({\bf x})$ and optimize it with respect to the parameters $\{\lambda_\nu\}$.

Maximizing with respect to $P({\bf x})$ give us
\begin{eqnarray}
0 &=& {{\partial {\cal S}\left[ P;\{\lambda_\nu\}\right]}\over{\partial P({\bf x})}} \nonumber\\
&=& {{\partial S\left [P\right ]}\over{\partial P({\bf x})}} - \sum_{\mu =0}^K   \lambda_{\mu} {{\partial  \langle f_{\mu}({\bf x})\rangle_P}\over{\partial P({\bf x})}}\nonumber\\
&=& - \ln P({\bf x}) - 1 - \sum_{\mu =0}^K \lambda_\mu f_\mu ({\bf x}),\\
\Rightarrow P({\bf x}) &=& {1\over {Z(\{\lambda_\nu\})}} \exp\left[ - \sum_{\mu = 1}^K \lambda_\mu f_\mu ({\bf x})\right] ,
\label{P_maxent}
\end{eqnarray}
where $Z(\{\lambda_\nu\}) = \exp(-\lambda_0 - 1)$.  Since optimizing with respect to $\lambda_0$ will enforce normalization of the distribution, we can write, explicitly,
 \begin{equation}
Z(\{\lambda_\nu\}) = \sum_{\bf x} \exp\left[ - \sum_{\mu = 1}^K \lambda_\mu f_\mu ({\bf x})\right] .
\label{ZA}
\end{equation}

Optimizing with respect to $\{\lambda_\nu\}$ gives us a set of $K$ simultaneous equations
\begin{eqnarray}
0 &=& {{\partial {\cal S}\left[ P;\{\lambda_\nu\}\right]}\over{\partial \lambda_\mu}}\nonumber\\
&=& \langle f_\mu ({\bf x})\rangle_{\rm exp} -  \langle f_\mu ({\bf x})\rangle_{P}\nonumber\\
\Rightarrow \langle f_\mu ({\bf x})\rangle_{\rm exp} &=& {1\over {Z(\{\lambda_\nu\})}} \sum_{\bf x}  f_\mu ({\bf x})\exp\left[ - \sum_{\mu = 1}^K \lambda_\mu f_\mu ({\bf x})\right]  \ .
\nonumber\\
&&
\label{multipliers}
\end{eqnarray}
Thus, when we optimize $\cal S$ with respect to the parameters $\{\lambda_\nu\}$ we are enforcing that the expectation values of the $\{ f_\mu({\bf x})\}$ agree with their experimental values, which is the starting point of the maximum entropy construction. Note also that, if we substitute Eq (\ref{P_maxent}) back into Eq (\ref{gen_entropy}), we obtain 
\begin{equation}
{\cal S}\left[ P;\{\lambda_\nu\}\right] = {\rm ln} Z(\{\lambda_\nu\}) + \sum_{\mu =0}^K  \lambda_{\mu} \langle f_{\mu}({\bf x})\rangle_{\rm exp}  \ ,
\label{gen_ent_max}
\end{equation}
which is minus the log probability, or likelihood, that the model generates the observed data. The optimal values of $\{\lambda_\nu\}$  correspond to minima of ${\cal S}$, as can be checked by considering the second derivatives. Therefore, the maximum entropy approach also corresponds to maximizing the likelihood that the model in Eq (\ref{ZA}) generates the observed data.

The maximum entropy distributions are familiar from statistical mechanics. Indeed we recall that a system in thermal equilibrium is described by a probability distribution that has the maximum possible entropy consistent with its average energy.  If the system has states described by a variable $\bf x$, and each state has an energy $E({\bf x})$, then this equilibrium distribution is 
\begin{equation}
P({\bf x}) = {1\over {Z(\beta )}} e^{-\beta E({\bf x})},
\label{boltzmann}
\end{equation}
where $\beta = 1/k_B T$ is the inverse temperature, and the partition function $Z(\beta )$ normalizes the distribution,
\begin{equation}
Z(\beta ) = \sum_{\bf x} e^{-\beta E({\bf x})}.
\end{equation}
In this view,  the  temperature is just a parameter we have to adjust so that the average value of the energy agrees with experiment.  The fact that equilibrium statistical mechanics is the prototype of maximum entropy models encourages us to think that the maximum entropy construction defines an effective ``energy'' for the system.  Comparing Eq's (\ref{P_maxent}) and (\ref{boltzmann}) gives us
\begin{equation}
E({\bf x}) = \sum_{\mu = 1}^K \lambda_\mu f_\mu ({\bf x}) ,
\end{equation}
and an effective temperature $k_B T = 1$.  This is a mathematical equivalence, not an analogy, and  means that we can carry over our intuition from decades of theoretical work on statistical physics.

In this paper, we discuss the case where the pairwise correlations $\langle {\vec s}_{\rm i} {\bf \cdot} {\vec s}_{\rm j}\rangle $ are measured experimentally.   Thus we can use the general maximum entropy formulation, identifying  ${\bf x} = \{{\vec s}_i\}$ and   $f_\mu({\bf x})={\vec s}_{\rm i}\cdot{\vec s}_{\rm j}$.  Since the quantities that will be measured refer to pairs, it is useful to set $\lambda_\mu=-J_{\rm ij}$, and we obtain Eq (\ref{maxent1}),
\begin{equation}
P(\{  {\vec s}_{\rm i}\}) = {1\over {Z(\{J_{\rm ij}\})}} \exp\left[ {1\over 2}\sum_{{\rm i}=1}^N \sum_{{\rm  j}=1}^N J_{\rm ij}  {\vec s}_{\rm i} {\bf \cdot} {\vec s}_{\rm j}\right] .
\label{maxent1A}
\end{equation}
As before,  the parameters $\{J_{\rm ij}\}$ must be adjusted so that  $\langle {\vec s}_{\rm i} {\bf \cdot} {\vec s}_{\rm j} \rangle_{P}=\langle {\vec s}_{\rm i} {\bf \cdot} {\vec s}_{\rm j} \rangle_{\rm exp}$.  

The model defined by  Eq (\ref{maxent1A}) is identical to a well known model for magnetism, the Heisenberg model. In that case, the model describes individual spins, which tend to mutually align according to the interactions  $J_{\rm ij}$. In this context, the effective energy is
\begin{equation}
E(\{ {\vec s}_{\rm i}\}) = - {1\over 2}\sum_{{\rm i}=1}^N \sum_{{\rm  j}=1}^N J_{\rm ij}  {\vec s}_{\rm i} {\bf \cdot} {\vec s}_{\rm j} .
\end{equation}
For $J_{\rm ij} > 0$, the energy is lowered when the vectors ${\vec s}_{\rm i}$ and ${\vec s}_{\rm j}$ are parallel.

\section{The spin wave approximation}
\label{App:spinwave}

The most demanding step in evaluating the probability distribution in Eq (\ref{maxent1A}) is the computation of the partition function  
\begin{equation}
Z(\{J_{\rm ij}\}) = \int d^N{\bf\vec s} \  \exp\left[  {1\over 2}\sum_{{\rm i}=1}^N \sum_{{\rm  j}=1}^N J_{\rm ij}  {\vec s}_{\rm i} {\bf \cdot} {\vec s}_{\rm j}\right] ,
\label{Heisenberg}
\end{equation}
where we recall that the $\{{\vec s}_i\}$ are  real, three dimensional vectors of unit length and $d^N{\bf\vec s}=\prod_{\rm i} d{\vec s}_{\rm i}$.

In presence of strong ordering, we can use the ``spin wave'' approximation to compute analytically the partition function of the Heisenberg model, Eq (\ref{Heisenberg}). Let us call ${\vec S}=(1/N) \sum_i{\vec s}_{\rm i}=S   {\vec n}$ the global order parameter, or polarization, measuring the degree of collective alignment, where $\vec n$ is a unit vector. Individual orientations can be rewritten in terms of a  longitudinal and a perpendicular component with respect to  ${\vec n}$,
\begin{equation}
{\vec s}_{\rm i} = s_{\rm i}^{\rm L} {\vec n} + {{\bf\vec \pi}}_{\rm i} \ ,
\label{longperp}
\end{equation}  
where, by construction, $\sum_{\rm i} s_{\rm i}^{\rm L}=S N$, ${\vec \pi}_{\rm i} \cdot {\vec n}=0$, and $\sum_{\rm i} {\vec \pi}_{\rm i}=0$.
The partition function then reads
\begin{widetext}
\begin{equation}
Z(\{J_{\rm ij}\}) = \int d^N{\bf s^L} \    d^N\vec{\mbox{\boldmath$\pi$} }  \left[  \prod_{\rm i} \delta\left( (s_{\rm i}^{\rm L})^2+|\vec{\pi}_{\rm i}|^2-1\right) \right]
 \delta\left(\sum_{\rm i} {\bf\vec\pi}_{\rm i}\right)
\exp\left[  {1\over 2}\sum_{{\rm i}=1}^N \sum_{{\rm  j}=1}^N J_{\rm ij}  \left ( s_{\rm i}^L s_{\rm j}^L +{\bf\vec \pi}_{\rm i} {\bf \cdot} {\bf\vec \pi}_{\rm j}\right) \right ]  \ ,
\end{equation}
\end{widetext}
where $ d^N{\bf s^L} = \prod_{\rm i} ds^L_{\rm i}$ and $d^N\vec{\mbox{\boldmath$\pi$} }=\prod_{\rm i} d{\vec \pi}_{\rm i}$. The delta functions implement  the constraint on the length of each vector  ${\vec s}_{\rm i}$ and the global constraint on the ${\vec \pi}_{\rm i}$.  Note that since the $\{ \vec\pi_{\rm i}\}$'s belong to the subspace perpendicular to $\vec n$, in Eq (\ref{spinwave}) there are only two independent degrees of freedom for each integration variable.

\begin{figure}[b]
\includegraphics[width = 0.6 \linewidth]{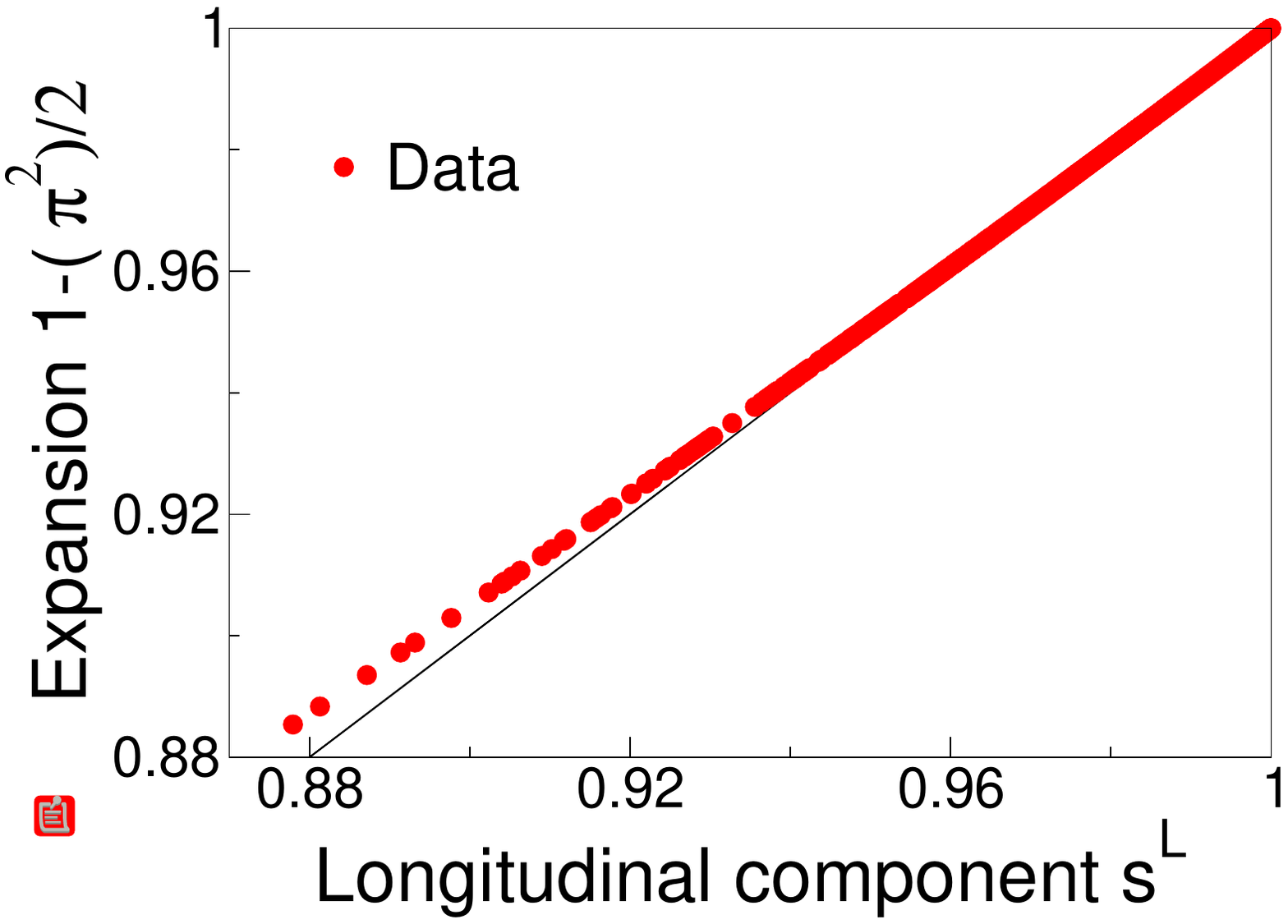}
\caption{Longitudinal components of the flight directions vs. prediction of the spin wave expansion, for all individuals in the  snapshot of Fig \ref{vectors}. The black line has slope 1. Note that $95\%$ of the birds have $s^{\rm L}_{\rm i} >0.94$, and lie well on the line.}
 \label{check_spinwave}
\end{figure}

If the system is highly polarized, $S\sim 1$ and  $|{\vec \pi}_{\rm i}| \ll 1$.  The constraint on the norm of the vectors can then be written as $s_{\rm i}^L\sim1- |\vec \pi_{\rm i}|^2/2$. Note that indeed flocks are very polarized groups and this expression is very well satisfied by the data (see Fig \ref{check_spinwave} and Table  \ref{table}). Using this expansion the longitudinal components can be integrated out easily. The partition function then becomes, to leading order in the ${\vec\pi}$'s, 
\begin{widetext}
\begin{equation}
Z(\{J_{\rm ij}\}) = \int d^N \vec{\mbox{\boldmath$\pi$} } \ \left[ \prod_{\rm i}  \ \frac{1}{ \sqrt{1-{|{\vec\pi}_{\rm i}|}^2}  }\right] \ \delta\left ( \sum_{\rm i} {\bf \vec \pi}_{\rm i} \right ) \exp{\left [ - \frac{1}{2} \sum_{\rm i,j =1}^N A_{\rm ij} {\vec \pi}_{\rm i}\cdot{\vec\pi}_{\rm j}+\frac{1}{2}\sum_{\rm i,j=1}^{N}J_{\rm ij}\right ]  }
\label{spinwave}
\end{equation}
\end{widetext}
with
\begin{equation}
A_{\rm ij}=\sum_k J_{\rm ik} \delta_{\rm ij} - J_{\rm ij}  \ .
\end{equation}
The product over $1/\sqrt{1-{|{\vec\pi}_{\rm i}|}^2}$ in   Eq (\ref{spinwave}) is the Jacobian coming from the integration over the ${s_{\rm i}}^{\rm L}$. This term gives rise to sub--leading contributions in the spin wave approximation, and we shall drop it. We have checked in our computations  that the corrections due to this term are indeed negligible.

The matrix $A$ is, by construction, a  positive semi--definite matrix.  We can find eigenvalues $a_{\rm k}$ and eigenvectors ${\bf w}^{\rm k}$ as usual through
\begin{equation}
\sum_{\rm j} A_{\rm ij} w_{\rm j}^{\rm k} = a_{\rm k} w_{\rm j}^{\rm k} .
\end{equation}
There is  one zero eigenvalue, $a_1=0$, corresponding to the constant eigenvector ${\vec w}^1=(1/\sqrt{N},1/\sqrt{N}, \cdots,1/\sqrt{N})$:
\begin{equation}
\sum_{\rm j} A_{\rm ij} w^1_{\rm i}= \frac{1}{\sqrt{N}}\sum_{\rm j} A_{\rm ij}=0 .
\end{equation}
The argument of the delta function in Eq (\ref{spinwave}) is related only to the projection of the $\{\vec \pi_{\rm i}\}$ onto this zero mode.  We note that  in a system with translation invariance, the eigenvectors are Fourier modes, or plane waves, and these are called spin waves in the theory of magnetism.   The zero eigenmode is related to the spontaneous breaking of symmetry when the flock chooses a consensus direction of flight---all directions ${\vec n}$ are equally probable, a priori, and hence have equal probability or energy, and the zero mode is the remanent of this symmetry; in physics this is the Goldstone mode.

We can now rewrite Eq (\ref{spinwave})  in the orthonormal basis defined by $\{{\vec w}^{\rm k}\}$:
\begin{widetext}
\begin{equation}
Z(\{J_{\rm ij}\})= \int d^N \vec{\mbox{\boldmath$\pi$}' } \ \delta\left ( {\bf\vec \pi'}_1\right )\exp{ \left [- \frac{1}{2}\sum_{{\rm k}=1}^N a_{\rm k} | {\bf\vec\pi'}_{\rm k} |^2
+\frac{1}{2}\sum_{\rm i,j=1}^{N}J_{\rm ij}\right ]  } ,
\label{spinwave1}
\end{equation}
\end{widetext}
where ${\vec\pi'_k=\sum_{\rm i} w^k_{\rm i} {\vec\pi}_{\rm i}}$.  This leads to 
\begin{equation}
\log Z(\{J_{\rm ij}\})= - \sum_{{\rm k} >  1} \log(a_{\rm k})+\frac{1}{2}\sum_{\rm {i,j}=1}^N J_{\rm ij}
\ ,
\label{Z_spinwave} 
\end{equation}
where we drop constant terms independent of $J_{\rm ij}$.

Let us now proceed, at a formal level, with the maximum entropy approach. The parameters $J_{\rm ij}$ are fixed by requiring that $\langle {\vec s}_{\rm i} {\bf \cdot} {\vec s}_{\rm j} \rangle_{P}=\langle {\vec s}_{\rm i} {\bf \cdot} {\vec s}_{\rm j} \rangle_{\rm exp}$. If we focus on the perpendicular part of the correlation, this implies
\begin{equation}
\langle {\bf\vec \pi}_{\rm i} {\bf \cdot} {\bf\vec \pi}_{\rm j} \rangle_{\rm exp}=2 \sum_{k>1} \frac{w^k_{\rm i}w^k_{\rm j}}{a_k} \ ,
\label{corr_ww}
\end{equation}
where the right hand side, the expectation value $\langle \vec{\pi}_{\rm i}\cdot\vec{\pi}_{\rm j}\rangle_P$, can be obtained from Eq (\ref{spinwave1}) using Gaussian integration rules, the factor 2 coming from the two independent degrees of freedom of each $\vec\pi_{\rm i}$.   
According to this equation, the matrix $A_{\rm ij}$---and therefore the interaction matrix $J_{\rm ij}$---is easily obtained by taking the inverse of the experimental perpendicular correlation function  (once we take away the zero mode due to symmetry).  But, to be invertible, the experimental correlation matrix must have $N-1$ nonzero eigenvalues. This can only be achieved by performing a huge number of experiments, i.e. evaluating the experimental average over a number of independent samples larger than the number of birds in the flock.  
As discussed  in the main text, the interaction network  in a flock changes continuously  in time, since individuals move and change their neighbors. But the average over many independent realizations of $\langle {\vec s}_{\rm i} {\bf \cdot} {\vec s}_{\rm j}\rangle $   would require birds to stay still  at some fixed positions, while updating and realigning their velocities, which is definitely not the case. In other terms, different experimental samples (i.e. snapshots) correspond  to different networks $J_{\rm ij}$ and cannot be averaged together. Thus, in our case, the maximum entropy model must be solved independently at each time step, for which we have only one experimental sample. Unfortunately, if we compute the correlation {\em matrix} from a single snapshot, it has rank two and cannot be inverted.  This motivates, as discussed in the text, the analysis of a more restricted problem, in which we know only the average local correlations $C_{\rm int}$ from Eq (\ref{corre_local}).

\section{Computation with free boundaries}
\label{App:freebound}

Let us now address more in details the reduced model, Eq (\ref{corre_local}) in main text, where each individual interacts with constant strength with its first $n_c$ neighbors. In this case the  $J_{\rm ij}$'s have a particularly simple form: 
\begin{equation}
J_{\rm ij}=J\ n_{\rm ij}
\end{equation}
with
\begin{equation}
n_{\rm ij} = \left \{ 
\begin{array}{ll}
1              &\ \ \ \ {\rm if} \ {\rm j} \in n_c^{\rm i}  \  {\rm and} \   \ {\rm i} \in n_c^{\rm j} \, ,\\ \\
\frac{1}{2} &\ \ \ \ {\rm if} \ {\rm j} \in n_c^{\rm i}  \  {\rm and} \    \ {\rm i} \notin n_c^{\rm j}\, , \ {\rm or\ vice\ versa, \ and} \\ \\
0               &\ \  \ \ {\rm otherwise}.
\end{array}
\right.
\label{nij}
\end{equation}
Here, $J$ indicates the strength of the interaction and $n_c^{\rm i}$ indicates the set of the first $n_c$ neighbors of bird $\rm i$.  Since we know the spatial coordinates of all the birds in the flock, once the parameter $n_c$ is fixed, we can compute all the neighborhoods and determine the matrix $n_{\rm ij}$. In this case, therefore,  $A_{\rm ij} = J \tilde A_{\rm ij}$, where $\tilde A$ depends only on the neighborhood relations.

 \begin{figure*}
\includegraphics[width = 0.35 \linewidth]{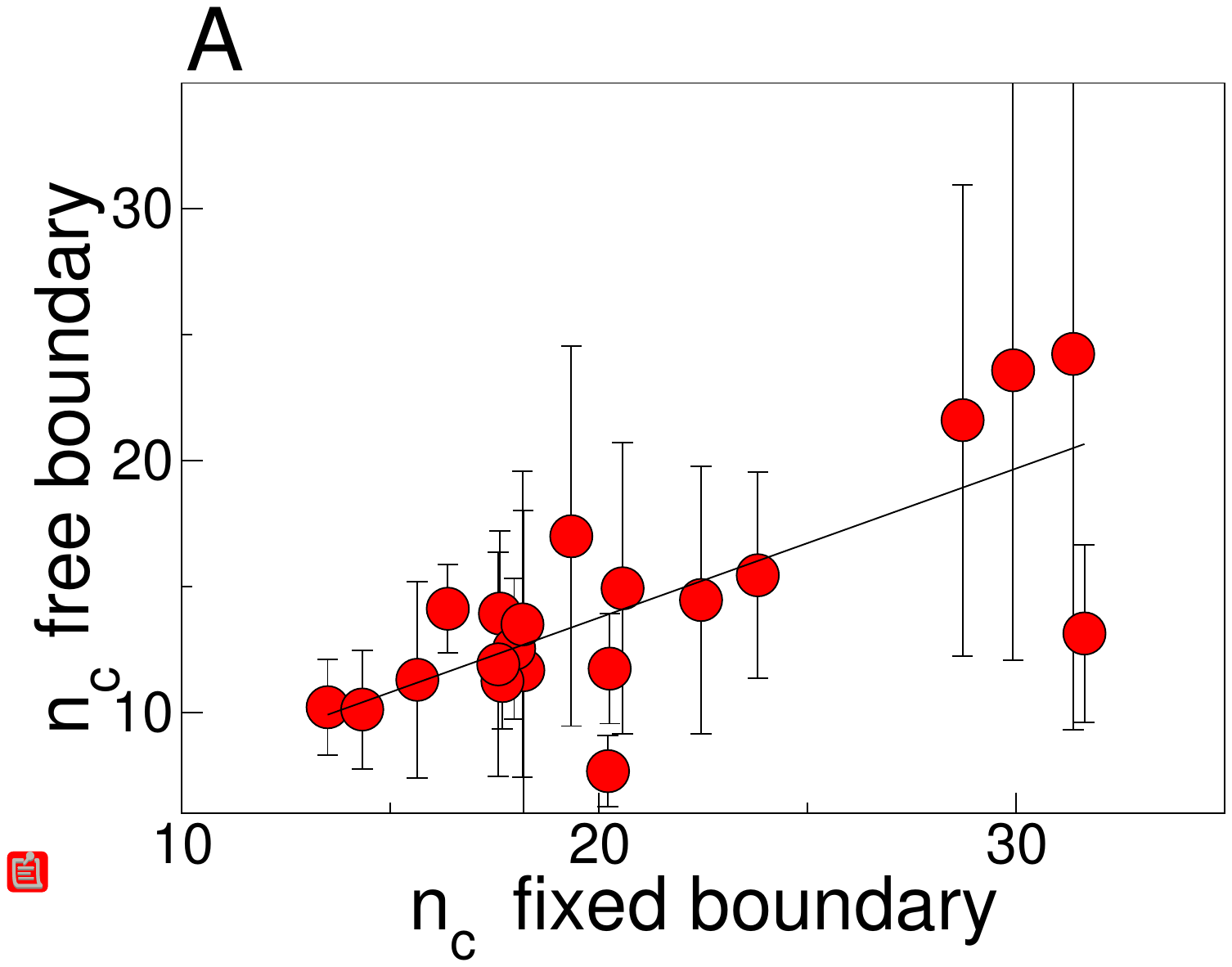}
\includegraphics[width = 0.35 \linewidth]{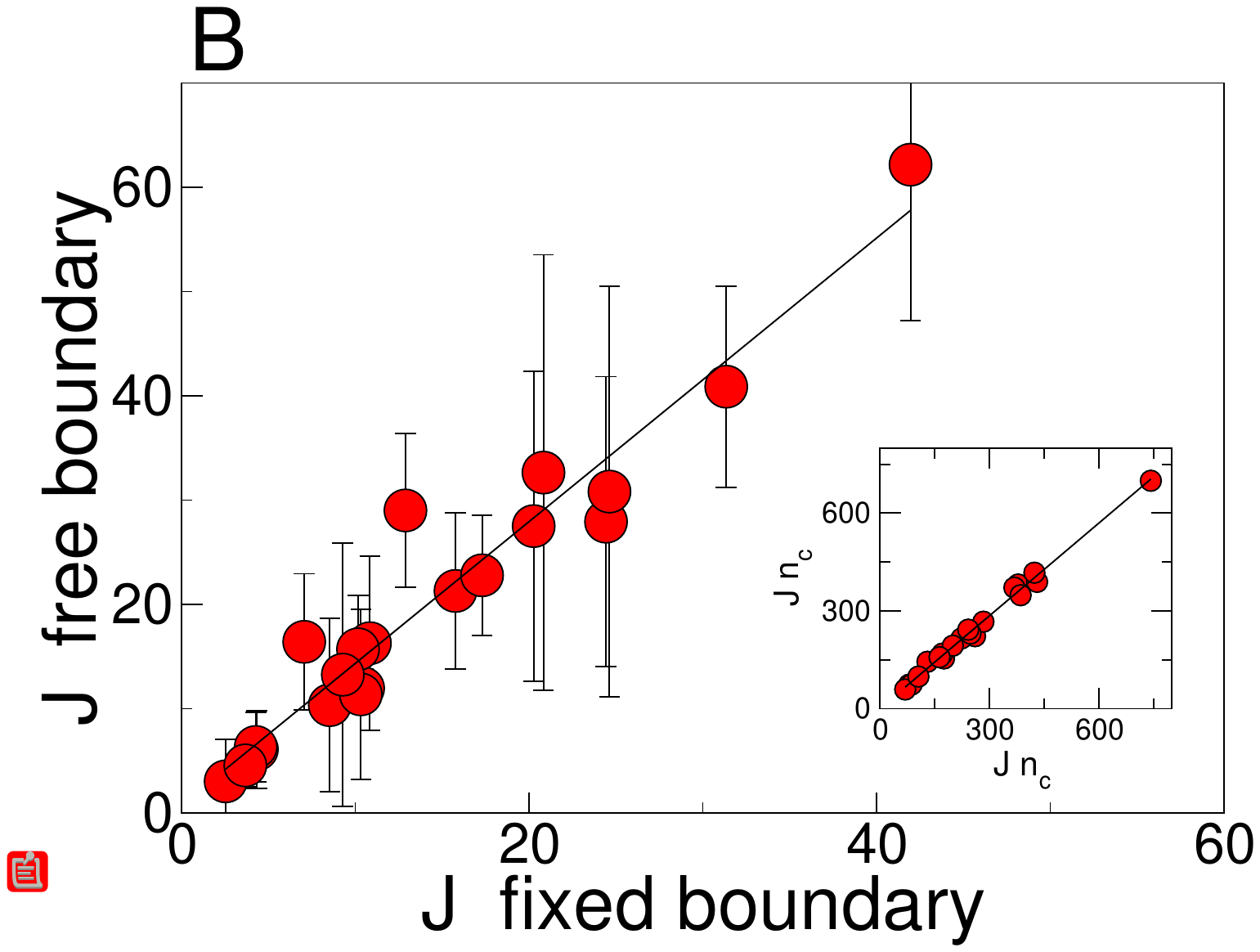}
\includegraphics[width = 0.35 \linewidth]{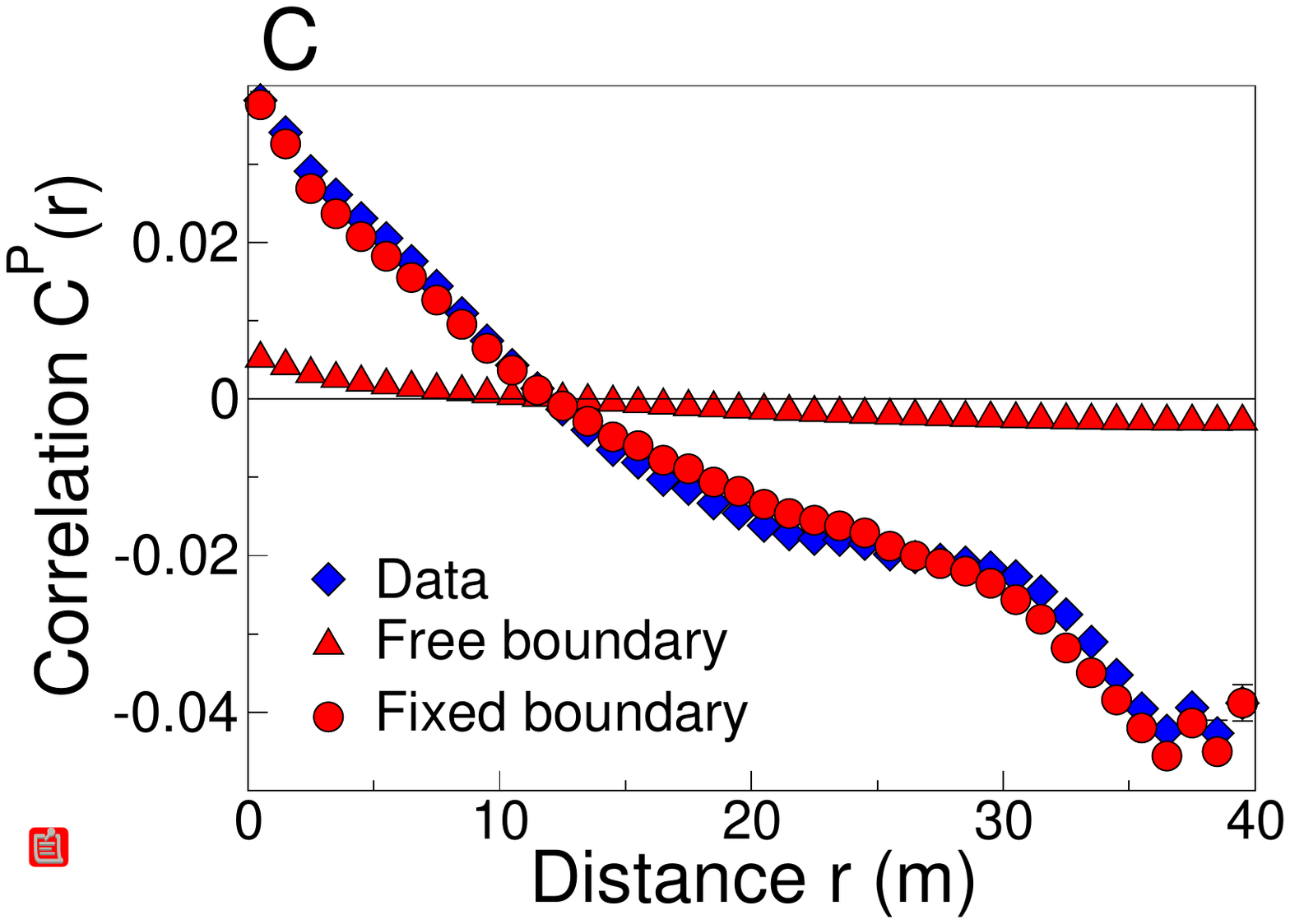}
\includegraphics[width = 0.35\linewidth]{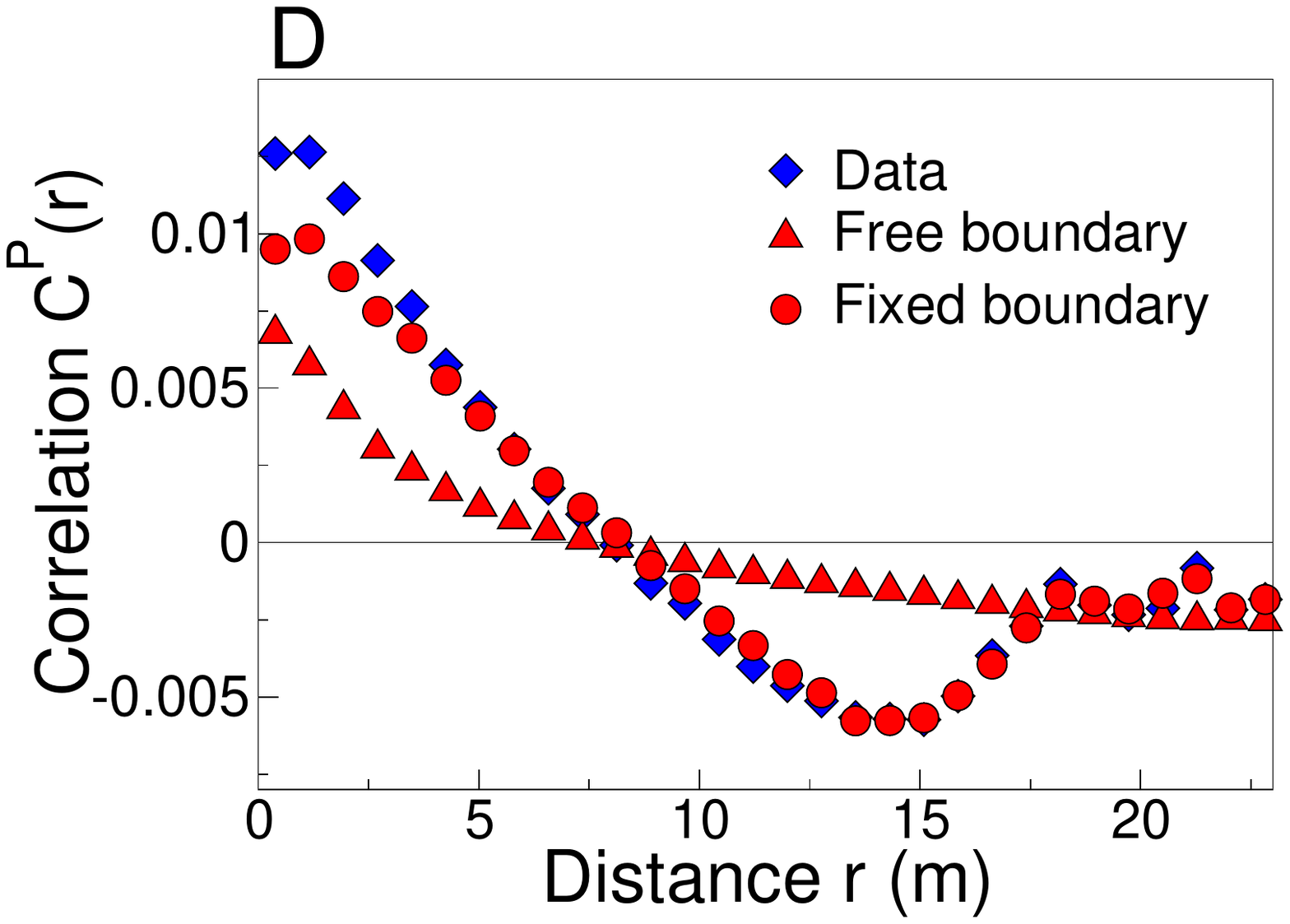}
\caption{Computation with free boundary conditions vs computation with fixed flight directions on the border. 
(A) Values of the parameter $n_c$ for all the flocking events; the black line is the linear regression. Error bars are standard deviations across multiple snapshots of the same flock. 
(B) Values of the parameter $J$ for all the flocking events. Inset: The product $J n_c$ computed with free boundary vs. $Jn_c$ computed with fixed boundary; now the slope is almost unity. 
(C) and (D) Perpendicular correlation as a function of distance for event 28--10 (as in Fig \ref{vectors};  $N=1246$ birds) and event 32--06 ($N=809$ birds). Different symbols correspond to  the correlation measured in experiments, the correlation computed with free boundary conditions and the one computed with fixed boundary conditions. Taking into account the flight directions of individuals on the border significantly improves the prediction for the correlation. }
 \label{border_vs_noborder}
\end{figure*}

Before proceeding with the full computation, let us briefly look at the simplest case, where we allow all the ${\vec \pi}_{\rm i}$'s to freely fluctuate according to Eq (\ref{spinwave1}). The result can be read directly from Eq (\ref{Z_spinwave}), giving
\begin{equation}
\log Z(J,n_c)= -\frac{1}{2}\sum_{k>1} \log(J \lambda_k)+\frac{NJn_c}{2}  \ ,
\label{Z_spinwavefree} 
\end{equation}
where the $\lambda_k$ are the eignevalues of $\tilde A$. Similarly, we can   compute the correlation functions,
\begin{eqnarray}
\langle {\vec \pi}_{\rm i}\cdot {\vec \pi}_{\rm j} \rangle &=& \frac{2}{J} \sum_{k>1} \frac{w^k_{\rm i}w^k_{\rm j}}{\lambda_k}
\nonumber \\
\langle s_{\rm i}^{\rm L}s_{\rm j}^{\rm L}\rangle&=&
1-\frac{1}{J}\sum_{k>1}  \frac{(w^k_{\rm i})^2+ (w^k_{\rm j})^2} {\lambda_k}
\label{correlation_noborder}
\end{eqnarray}
where $\lambda_k$ and ${\bf w}^k$ are, again, the eigenvalues and the eigenvectors of the matrix $\tilde A$ and depend only on $n_c$.

To build the maximum entropy model, we need to find the appropriate values for $J$ and $n_c$. As discussed in Appendix \ref{App:maxent}, this amounts to maximizing the log--likelihood of the experimental data given the model.   This can be written simply as
\begin{equation}
{\bigg\langle} \log P\left ( \{{\vec s}_{\rm i}\} \right ){\bigg\rangle}_{\rm exp}  = - \log Z(J,n_c)+\frac{1}{2} J N n_c C_{\rm int}^{\rm exp} \ ,
\label{likelihood_noborder}
\end{equation}
where, following the notation in the main text,
 \begin{equation}
 C_{\rm int}= \frac{1}{N}\sum_{\rm i}\frac{1}{n_c}\sum_{{\rm j}\in n_c^i} \langle {\vec s}_{\rm i}\cdot {\vec s}_{\rm j}\rangle
\end{equation}
 indicates the degree of correlation up to the interaction range $n_c$. 
 
Maximizing with respect to $J$, which corresponds to setting $C_{\rm int}(J,n_c) = C_{\rm int}^{\rm exp}$,  gives immediately
 \begin{equation}
\frac{1}{J}=\frac{n_c}{2} \left ( 1-C_{\rm int}^{\rm exp}\right ) \ .
\label{J_noborder}
\end{equation}
This equation provides an explicit relationship between $J$ and $n_c$. Substituting into Eq (\ref{likelihood_noborder}), the likelihood becomes a function of $n_c$ only, and its maximum can be  found numerically. 

The values of $J$ and $n_c$, obtained from this computation with free boundaries, are displayed in Fig \ref{border_vs_noborder},  for all the flocking events we analyzed. They are strongly correlated to what we find with fixed boundary conditions (see next section): the value of $n_c$ is slightly smaller, the value of $J$ slightly larger, but the product $Jn_c$ is approximately the same. On the contrary, the prediction for the perpendicular correlation as a function of distance (Fig \ref{border_vs_noborder}C and D) is less satisfactory: while the correlation length is correctly reproduced, the decay of the correlation with distance is significantly faster. Besides, the value of the perpendicular correlation near $r=0$  looks much smaller than the experimental value. To better understand this point we note that
\begin{equation}
C_{\rm int}=C_{\rm int}^{\rm P} + S + \left ( 1 - \frac{1}{N}\sum_{\rm i} \frac{1}{n_c} \sum_{\rm j \in {n_c}^{\rm i}} \langle |{\vec \pi}_{\rm j}|^2\rangle \right )  \ ,
\label{noborder_split}
\end{equation}
where, we recall, $S$ is the polarization.  The first term in this decomposition of $C_{\rm int}$  represents the perpendicular part of the correlation up to scale $n_c$, while the last term is a `local' polarization getting contributions only from individuals on a scale $n_c$. The maximum entropy model,  by construction, reproduces correctly the experimental value of $C_{\rm int}$. What happens in the computation with free boundaries  is that the model underestimates the contribution on short scales ($n<n_c$, corresponding to spatial scales of a few meters) from the perpendicular part of the correlation, and compensates by overestimating the polarization. The effect is more or less strong in different flocks, as seen in Figs \ref{border_vs_noborder}C and D.

As discussed in the main text, there are good reasons to think that birds on the edge of the flock should be described differently from those in the bulk;   Fig \ref{border_vs_noborder}C is evidence that if we ignore these differences we really do fail to predict correctly the correlation structure of the flock as a whole.

\section{Computation with fixed boundaries}
\label{App:fixedbound}

To improve our approach, we need to consider more appropriate boundary conditions. As discussed in the main text, birds on the border of the flock are likely to behave differently from birds in the interior of the flock. This occurs because they experience a different kind of neighborhood, part of the space around them being devoid of neighbors. Besides, these birds are continuously exposed to external stimuli and their dynamics may be strongly influenced by environmental factors (approaching predators, obstacles, nearby roosts, ...). Thus, modelling birds on the border might require taking into account other ingredients than the interactions between individuals.   Rather than trying to making a model of these (largely unknown) factors, we can take the velocities of these border birds as given, and ask that our model of interactions predict the propagation of order throughout the bulk of the flock.

If we consider the flight directions of birds on the border as given, the computation of the partition function becomes more complicated. The starting point is analogous to Eq (\ref{Heisenberg}), but the integration must be performed with respect to internal variables only. It is then convenient to separate, in the exponent of Eq (\ref{Heisenberg}), contributions coming from internal and external birds. Let us call ${\cal I}$ and ${\cal B}$ the subsets of internal and border individuals, respectively. Then, in the spin wave approximation, we find an expression similar to Eq (\ref{spinwave}):
\begin{widetext}
\begin{equation}
Z(\{J_{\rm ij}\};{\cal B})
= \int d^{\cal I} \vec{\mbox{\boldmath$\pi$} } \ \ \delta\left (\sum_{\rm i} {\bf\vec\pi}_{\rm i} \right) 
\exp\left[ - \frac{1}{2} \sum_{{\rm i,j}\in{\cal I}} A_{\rm ij} {\vec \pi}_{\rm i}\cdot{\vec\pi}_{\rm j}+\sum_{{\rm i} \in {\cal I}} {\vec\pi}_{\rm i}\cdot  {\vec h}_{\rm i}  + \frac{1}{2}\sum_{{\rm i,j}\in{\cal I}} J_{\rm ij} 
 +\frac{1}{2}\sum_{{\rm i} \in{\cal I}}{h}^{\rm L}_{\rm i}+\frac{1}{2}\sum_{{\rm i,j}\in{\cal B}}J_{\rm ij}{\vec s}_{\rm i}\cdot{\vec s}_{\rm j} \right ] \ ,
\label{Z_border}
\end{equation}
\end{widetext}
where 
\begin{eqnarray}
{\vec h}_{\rm i} &= & \sum_{{\rm l} \in {\cal B}} J_{\rm il} {\vec s}_{\rm l}
= \sum_{{\rm l} \in {\cal B}} J_{\rm il} \left( 
s_{\rm l}^{\rm L} {\vec n}  + {\vec \pi}_{\rm l} 
\right) = h_{\rm i}^{\rm L}{\vec n}+{\vec h}_{\rm i}^{\rm P}\nonumber \\
&&\\
A_{\rm ij} &= &\delta_{\rm ij} \left(\sum_{k\in{\cal I}} J_{\rm ik}+h_{\rm i}^{L}\right)-J_{\rm ij}\ \ \ \ \ \ {\rm i,j}\in{\cal I}
\label{def_border}
\end{eqnarray}
Here ${\vec h}_{\rm i}$ is a `field' describing the influence of birds on the border on internal bird $\rm i$. The effect of this field is to align bird $\rm i$ with the border birds that are within its direct interaction neighborhood $n^{\rm i}_c$.  Thus, when $n_c$ is small, this field only acts on individuals close to the border, while it is zero well inside the flock.   We also note that, as compared to Eq (\ref{spinwave}), the matrix $A$ is now defined only for internal birds and gets an additional diagonal contribution coming from individuals on the border. As a result, $A$ has no longer a zero mode. From a conceptual point of view,  when we fix the direction of motion of birds on the border, not all directions in the bulk are  a priori equivalent; rather, the boundary conditions explicitly break the symmetry. From a computational point of view, this implies that we cannot express in a simple way the constraint on the $\{{\vec \pi}_{\rm i}\}$'s as we did in the case of a free boundary. 

\begin{widetext}
To deal with the constraint, it is convenient to use an integral representation of the delta function
\begin{equation}
\delta\left( \sum_{\rm i} {\vec \pi}_{\rm i}\right) = \int {{d{\vec z}}\over{(2\pi)^2}}  \exp\left [ i  {\vec z} \cdot\sum_{\rm i}{\bf\vec\pi}_{\rm i}\right ] .
\end{equation}
Substituting into Eq (\ref{Z_border}), we obtain
\begin{equation}
Z(\{J_{\rm ij}\};{\cal B}) = 
\int {{d{\vec z}}\over{(2\pi)^2}} \int d^{\cal I} \vec{\mbox{\boldmath$\pi$} } \ \ \exp\left [ - \frac{1}{2}\sum_{{\rm i,j}\in {\cal I}} A_{\rm ij}{\vec\pi}_{\rm i}\cdot{\vec\pi}_{\rm j} 
+\sum_{{\rm i}\in{\cal I}} {\vec\pi}_{\rm i}\cdot \left({\vec h}^{\rm P}_{\rm i}+ i {\vec z} \right )
 +{ i} {\vec z}\cdot\sum_{{\rm l}\in {\cal B}}{\vec \pi}_{\rm l} + G({\cal B})\right ] ,
\label{Z_border_com}
\end{equation}
where $G({\cal B})$ is a function of boundary variables only. We notice that all the integrals are Gaussian, and we obtain, finally,
\begin{eqnarray}
\ln Z\left ( \{J_{\rm ij}\};{\cal B}\right ) &=& \frac{1}{2}\sum_{{\rm ij}\in {\cal I}} (A^{-1})_{\rm ij} \ {\vec h}^{\rm P}_{\rm i}\cdot {\vec h}^{\rm P}_{\rm j} -\ln \det  (A)  -\ln\left[ \sum_{{\rm ij}\in{\cal I}} (A^{-1})_{\rm ij}\right]
-\frac{1}{2}\frac{ \left [ \sum_{\rm l\in{\cal B}} {\vec\pi}_{\rm l}+\sum_{\rm ij\in{\cal I}} (A^{-1})_{\rm ij} {\vec h}^{\rm P}_{\rm j}\right ]^2 }{\sum_{\rm ij\in{\cal I}} (A^{-1})_{\rm ij}}
 \nonumber\\
 && \,\,\,\,\,\,\,\,\,\,\,\,\,\,\,\,\,\,\, +\frac{1}{2} \sum_{{\rm ij}\in {\cal I}} J_{\rm ij}  +\sum_{{\rm i}\in{\cal I}} h^{\rm L}_{\rm i} + \sum_{{\rm lm}\in{\cal B}}J_{\rm lm}{\vec s}_{\rm l}\cdot {\vec s}_{\rm m} ,
\label{Z_border2}
\end{eqnarray}
where $G({\cal B})$ is written explicitly. Recall that  the matrix $A$ is only defined on internal individuals and hence  the number of eigenvalues that contribute to the computation of $\det (A)$ is given by the number of internal birds.
In the same way, we can easily compute correlation functions. We find
\begin{equation}
\langle{\vec \pi}_{\rm i}\rangle = \sum_{\rm j\in{\cal I}} (A^{-1})_{\rm ij} {\vec h}^{\rm P}_{\rm i}- 
\frac{ \sum_{\rm j\in{\cal I}} (A^{-1})_{\rm ij}    
}
{\sum_{\rm kj\in{\cal I}} (A^{-1})_{\rm kj}}
\left[ \sum_{\rm l\in{\cal B}} {\vec\pi}_{\rm l} +\sum_{\rm kj\in{\cal I}} (A^{-1})_{\rm kj} {\vec h}^{\rm P}_{\rm j}\right] ,
\label{pimed_border}
\end{equation}
and
\begin{equation}
\langle {\vec \pi}_{\rm i}\cdot{\vec \pi}_{\rm j}\rangle  =  \langle{\vec \pi}_{\rm i}\rangle \cdot \langle{\vec \pi}_{\rm j}\rangle 
+ 2 \left [ (A^{-1})_{\rm ij} -\frac{ \sum_{\rm kn\in{\cal I}} (A^{-1})_{\rm ik}(A^{-1})_{\rm nj}} {\sum_{\rm kn\in{\cal I}}(A^{-1})_{\rm kn}}
\right ] .
\label{corre_border}
\end{equation}
\end{widetext}

At this point, to solve the maximum entropy model for the reduced case, we simply substitute the parametrization $J_{\rm ij}=J n_{\rm ij}$. The log--likelihood takes the form
\begin{equation}
{\bigg \langle} \log P\left(\{{\vec s}_{\rm i}\}\right){\bigg \rangle}_{\rm exp} =- \log Z(J,n_c;{\cal B}) +\frac{1}{2}J n_c N C^{\rm exp}_{\rm int} \ .
\label{loglikelihood_border}
\end{equation}
with $Z(J,n_c;{\cal B})$ as in Eq (\ref{Z_border}). To find the optimal value for the parameters $J$ and $n_c$ we need to maximize the likelihood. Maximization with respect to $J$ again is equivalent to matching the predicted correlations to the experimental ones,  $C_{\rm int}(J,nc;{\cal B})=C^{\rm exp}_{\rm int}$. This equation is represented graphically in Fig \ref{fit}A in the main text.  It is worth noting that, as in the case with free boundary conditions, it is possible to solve this equation analytically.   We can define
\begin{eqnarray}
{\tilde A} &=& A/J , \\
 \tilde{\vec h}_{\rm i} &=& {\vec h}_{\rm i}/J,
\end{eqnarray}
both of which are independent of $J$, and then, after some algebra, we obtain
\begin{widetext}
\begin{equation}
\frac{(N_{\rm in}-1)}{J}=
\frac{1}{2}\sum_{{\rm ij}\in {\cal I}} ({\tilde A}^{-1})_{\rm ij}  \tilde{\vec h}^{\rm P}_{\rm i} \cdot  \tilde{\vec h}^{\rm P}_{\rm j}
-\frac{1}{2} \frac{ \left [ \sum_{\rm l\in{\cal B}} {\vec\pi}_{\rm l}+\sum_{\rm ij\in{\cal I}} ({\tilde A}^{-1})_{\rm ij} \tilde{\vec h}^{\rm P}_{\rm j}\right ]^2 }{\sum_{\rm ij\in{\cal I}} ({\tilde A}^{-1})_{\rm ij}}+\sum_{{\rm i}\in{\cal I}}  {\tilde h}^{\rm L}_{\rm i}  +
\sum_{{\rm lm}\in{\cal B}}n_{\rm lm}{\vec s}_{\rm l}\cdot {\vec s}_{\rm m}+\frac{N n_c }{2}(1-C^{\rm exp}_{\rm int})
\label{final_J}
\end{equation}
\end{widetext}
where $N_{\rm in}$ is the number of internal birds. Note that the right hand side is a function of $n_c$ only, so we have an expression for  $J(n_c;{\cal B})$. Substituting back into Eq (\ref{loglikelihood_border}) we get the likelihood as a function of $n_c$ only. Maximization can be performed numerically, as shown in Fig \ref{fit}B in the main text.

Values of $J$ and $n_c$ for all flocks are collected in Figs \ref{fit} in the main text and  in Fig \ref{border_vs_noborder}. In this figure, we see the improvement in the prediction of the  correlation function $C(r)$ that comes with fixed  boundary conditions.

\section{A global model}
\label{App:global}

\begin{figure}[bht]
\includegraphics[width = 0.48 \linewidth]{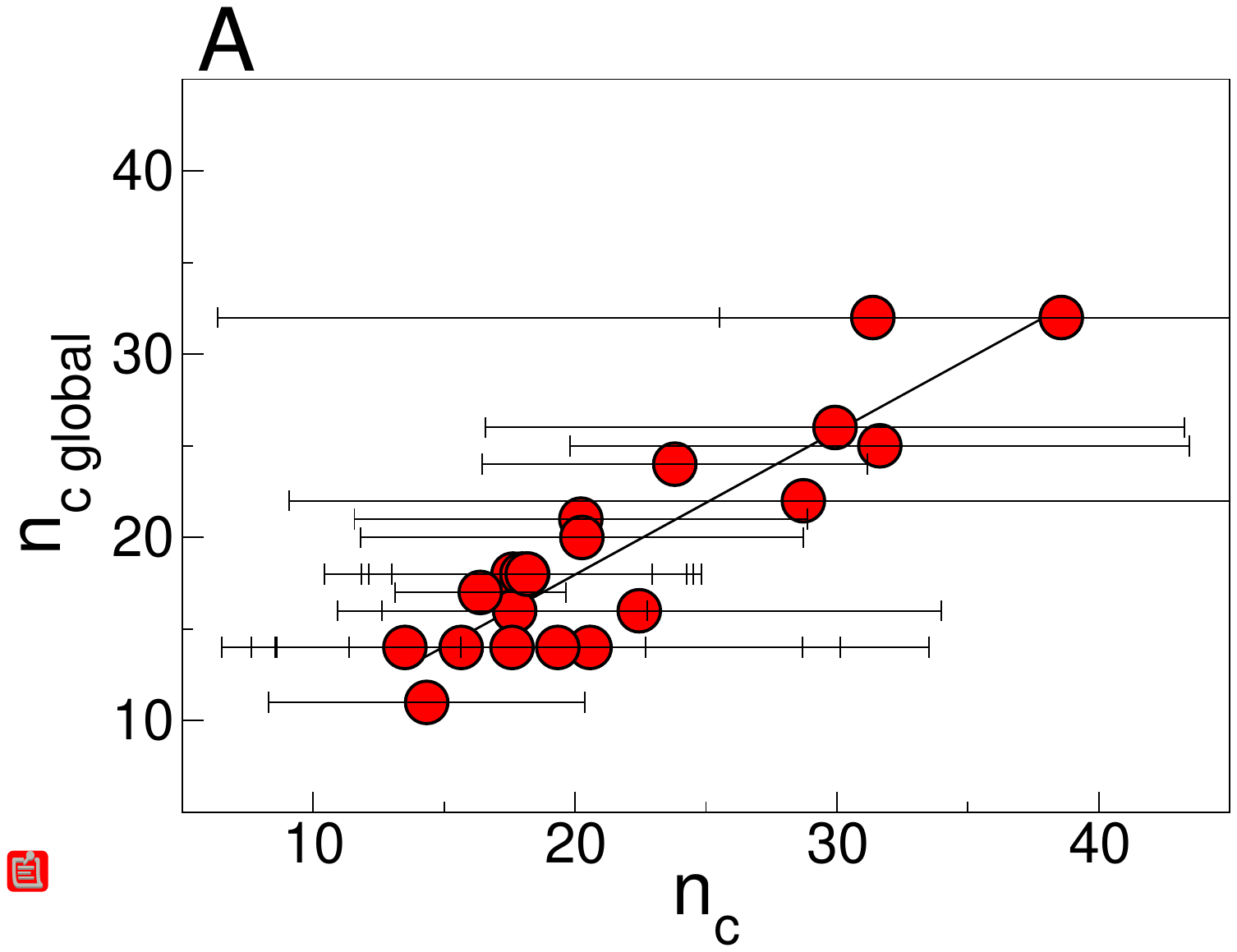}
\includegraphics[width = 0.48 \linewidth]{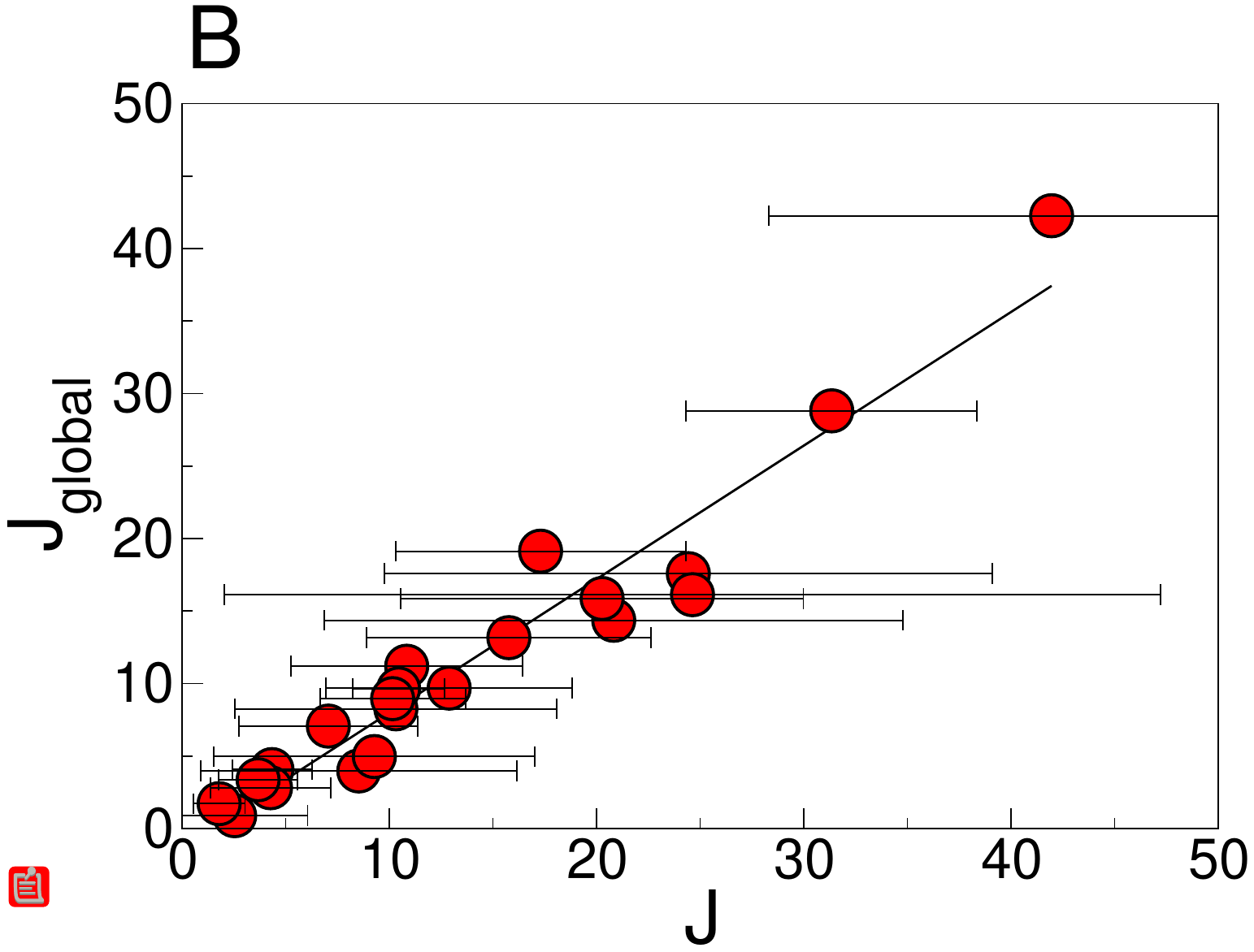}
\caption{Global models of flocking events.  
(A)  Values of the neighborhood size $n_c$ inferred from maximizing the log--likelihood averaged over snapshots, plotted vs. the mean values obtained from maximizing log--likelihood in individual snapshots.   Error bars represent the standard deviation over snapshots for each flock.  Black line has a slope 0.78.
(B) As in (A), but for the interaction strength $J$; the black line has slope 0.92.
\label{global_vs_snapshot}}
\end{figure}

Given a flock of birds,  so far we have solved the maximum entropy model for each individual snapshot independently, and then we have averaged the inferred values of the parameters $J$ and $n_c$ over all the snapshots. This is the most general procedure we can use, consistent with the dynamical nature of the interaction network. The inferred values of $J$ and $n_c$ fluctuate from snapshot to snapshot, due to several factors. It is possible that birds slightly adjust interaction strength and range during time, but there are other noisy contributions that might increase the fluctuations. The flocks we analyzed are finite groups, ranging from a few hundreds to a few thousands individuals, and we therefore expect finite size effects. The algorithmic procedure to reconstruct positions and velocities of individual birds in the flock is very efficient but not perfect, and there are fluctuations across snapshots in the number of reconstructed individuals; see Refs \cite{ballerini+al_08b,cavagna+al_09} for details on the 3D reconstructions. Finally, the log--likelihood can be very flat in the region of the maximum: in this case even small fluctuations can cause the value of the maximum to jump from a value of $n_c$ to another one quite different. Averaging $n_c$ and $J$ over the snapshots, we get rid of these fluctuations. Alternatively, we can assume from the start that, given a flock, there is a unique value of $n_c $ and $J$ through time. In this case, the log--likelihood of each snapshot is a function of the {\it same} $J$ and $n_c$ and we need to optimize the global likelihood corresponding to all the snapshots, and not each one independently. In other terms, we first compute the average of the log--likelihood over the snapshots at $J$ and $n_c$ fixed, and then we maximize with respect to the two parameters. Note that we are inverting the procedure described in the previous sections, where, on the contrary, we first maximize each individual snapshot with respect to $J$ and $n_c$ and then we take the average over all the snapshots of the optimal parameters.
The computation of the average log--likelihood can be easily done starting from the equations for the single snapshot. Let us denote, for future convenience, by
\begin{equation}
\phi_\alpha(J,n_c)=- \log Z(J,n_c;{\cal B}_\alpha) +\frac{1}{2}J n_c N C^{\rm exp}_{\rm int ,\alpha} 
\end{equation}
the log--likelihood of the snapshot $\alpha$ with parameters $J$ and $n_c$ (see Eq (\ref{loglikelihood_border})). Then, the average log--likelihood for all the snapshots is
\begin{equation}
\Phi_{\rm global}(J,n_c) = \frac{1}{N_{\rm snap}}\sum_{\alpha} \phi_{\alpha}(J,n_c) \ ,
\label{likelihood_global}
\end{equation}
where $N_{\rm snap}$ is the number of snapshots available for that flock. At this point, we need to maximize $\Phi_{\rm global}$ over $J$ and $n_c$. The maximization with respect to $J$ leads, once again, to an explicit expression for the optimal $J$, that we shall call $J_{\rm global}$, as a function of $n_c$:
\begin{equation}
\frac{1}{J_{\rm global}}=\frac{1}{N_{\rm snap}} \sum_{\alpha} \frac{ (N_{\rm in}^\alpha-1) } { (N_{\rm in}^{\rm global}-1) } \frac{1}{J_{\alpha}(n_c)}
\end{equation}
where $J_{\alpha}(n_c)$ is the optimal value of $J$ for the snapshot $\alpha$, as above, $N_{\rm in}^\alpha$ is the number of internal individuals in the snapshot $\alpha$,  and $N_{\rm in}^{\rm global}$ is the average over snapshots. Substituting $J_{\rm global}(n_c)$ back in Eq (\ref{likelihood_global}) we get an expression, which is a function of $n_c$ only. The likelihood can then be maximized numerically with respect to $n_c$. The values  $n_{c , {\rm global}}$ and $J_{\rm global}$ obtained in this way are plotted in Fig \ref{global_vs_snapshot}, where they are compared to the values inferred with the more general procedure (optimizing each snapshot independently and then averaging). There is a very strong correlation with slope close to one. This represents a strong consistency check on the inference procedure.

\section{A model for order propagation}
\label{App:model}

\begin{figure}[bht]
\includegraphics[width = 0.48 \linewidth]{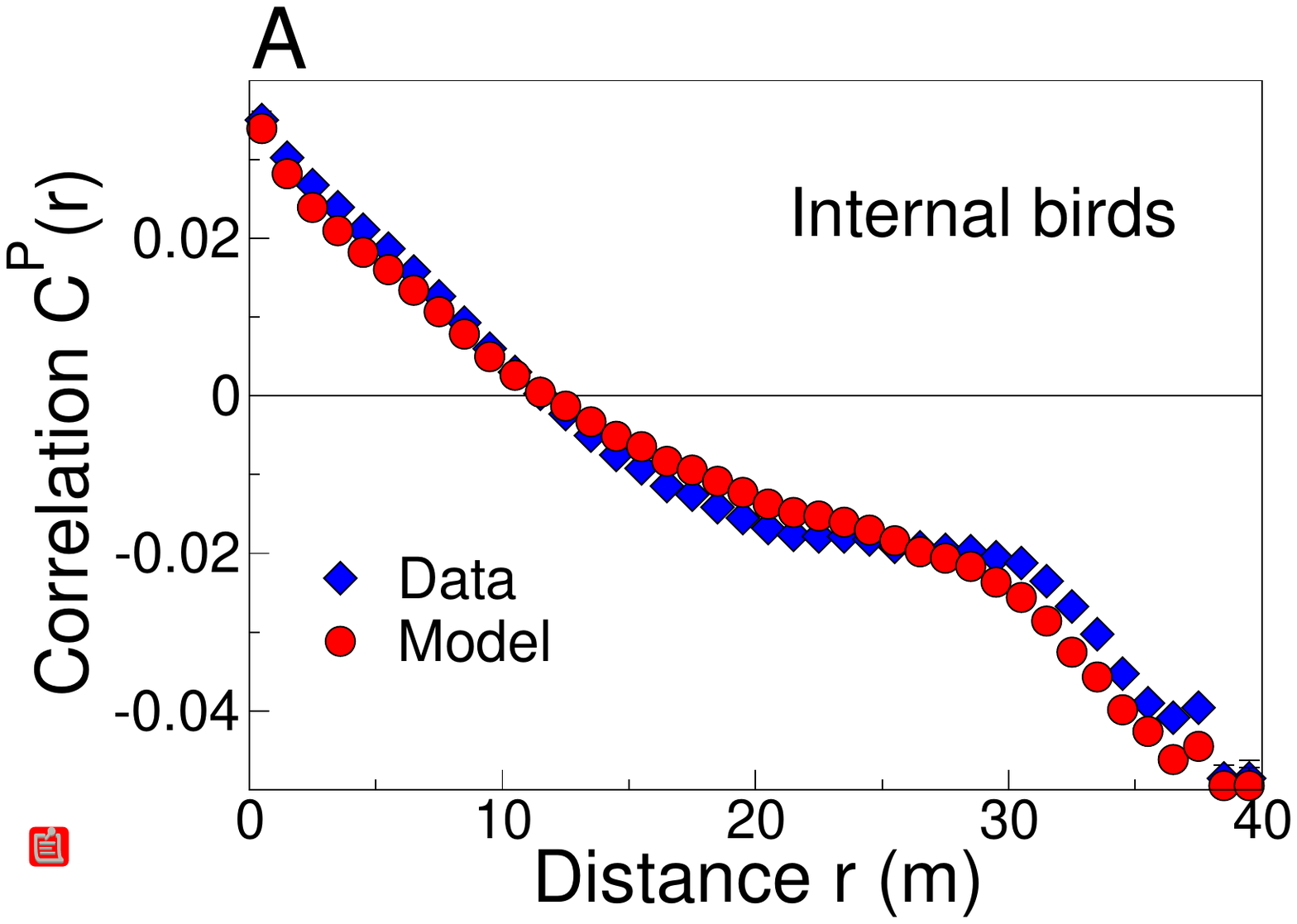}
\includegraphics[width = 0.48 \linewidth]{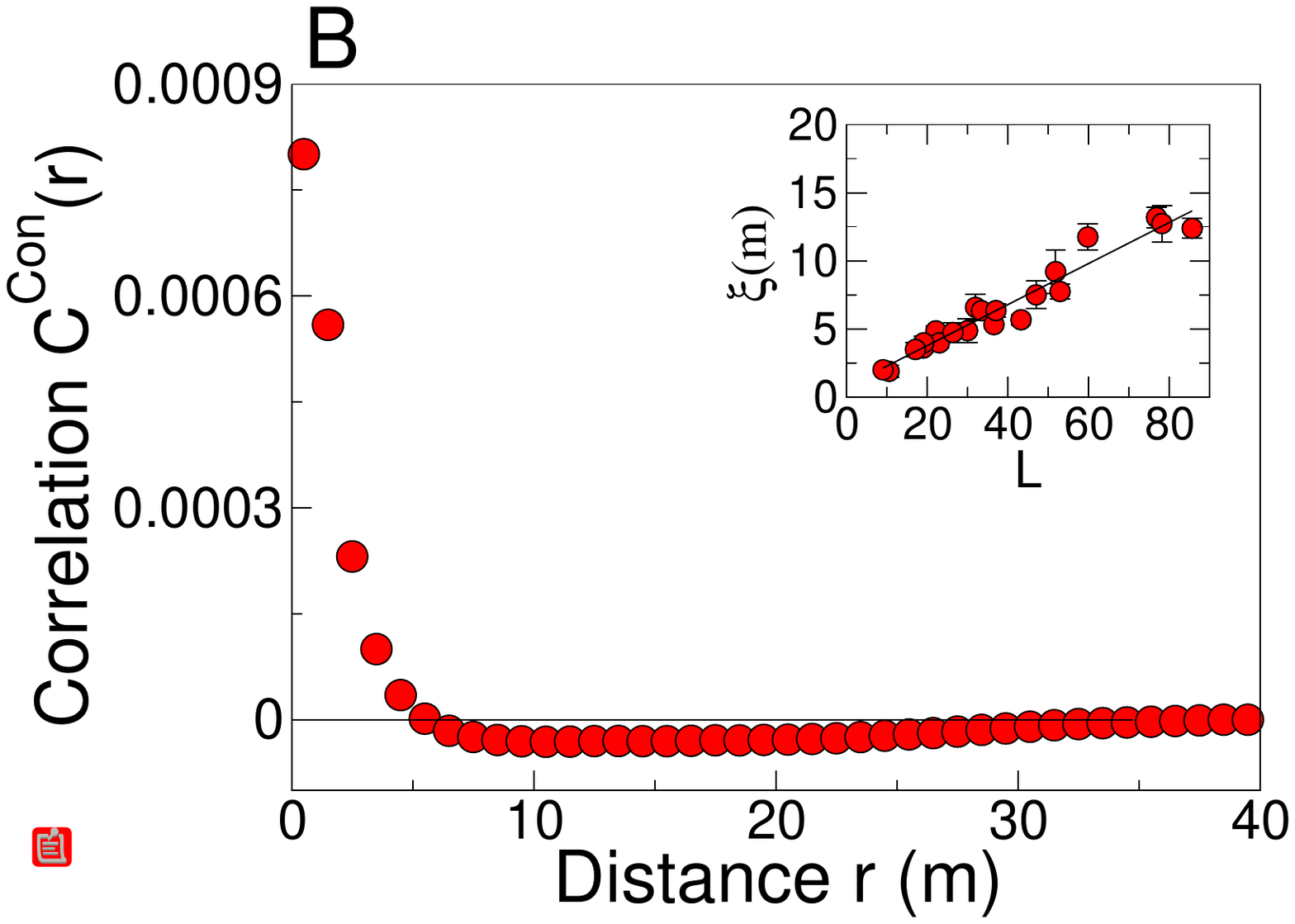}
\caption{Correlations in the interior of the flock.
(A) Perpendicular component of the two--point correlation function (as in Fig \ref{correlations}A) for internal birds only, as a function of distance. 
(B) Connected correlation function predicted in the model, as a function of distance. Inset: correlation length vs. flock size, for all the flocks that we analyzed.}
 \label{fig-corre_border}
\end{figure}

The maximum entropy model with fixed flight directions on the border gives excellent predictions for two--point and higher order correlation functions; see Figs \ref{correlations}, \ref{border_vs_noborder} and \ref{fig-corre_border}.  In addition, it allows to infer---up to a calibration factor---the microscopic interactions in a numerical model of self--propelled particles. We can conclude that this model indeed offers a very good statistical description of the flight directions of individuals in a flock. Let us then look back at the model, and try to understand the kind of system that the model describes.

We recall that, in this version of the model, we take as fixed the flight directions of the individuals on the border. Therefore, the model does not aim at predicting properties of border individuals, which, as we noted, may depend on  factors other than mutual interactions. Rather, the model focuses on internal individuals and how ordering flows through the flock. The state of the birds  on the border generate a `field'  (${\vec h}_{\rm i}$) on internal individuals, but   this field  is nonzero only for individuals interacting directly interacting with birds on the boundary (i.e. when $J_{\rm ij}=J n_{\rm ij}\ne 0)$. For the values of $n_c$ retrieved by the model ($n_c \sim 20$), this is only a small shell close to the border: all individuals well inside the flock, on the contrary, do not experience any direct influence from the border.

Still, if the model does describe what happens in a real flock, it must predict collective coherence: {\it all} flight directions must strongly align and internal individuals must behave very much in unison with their exterior companions. Does the model reproduce this behaviour?  If so, what is the mechanism leading to this kind of ordering? How do individuals on the border transmit information about their flight directions to distant individuals with whom they do not interact  directly?

The formal answer to these questions can be read in Eq's (\ref{pimed_border}) and (\ref{corre_border}). The first equation indicates that the model predicts a well defined perpendicular component of the flight direction $\langle {\vec \pi}_{\rm i}\rangle$ for each internal individual $\rm i$.  Surprisingly, these perpendicular components agree remarkably well with the ones measured experimentally (see Fig \ref{correlations}D in main text), not only  for birds close to boundary, but also well inside the group. The second equation provides a prediction for the correlation function. Visualization of these correlations as a function of distance shows that these predictions also are very good.  We note that, since the longitudinal component of the flight direction is given by $\langle s_{\rm i}^{\rm L}\rangle =1-0.5 \langle |{\vec\pi}_{\rm i}|^2\rangle $, if we are getting the perpendicular components of the velocity right we must also be getting the longitudinal components right.  Equations (\ref{pimed_border}) and (\ref{corre_border}) therefore provide correct predictions of the full flight directions for all individuals in the flock.

The mechanism through which such ordering occurs, is the presence of long ranged correlations in the system. This can be seen more easily rewriting the equations in the following way:
\begin{eqnarray}
\langle{\vec \pi}_{\rm i}\rangle & =& \sum_{\rm j\in{\cal I}} C^{\rm con}_{\rm ij} {\vec h}^{\rm P}_{\rm j}-
\frac{ \sum_{\rm j\in{\cal I}} (A^{-1})_{\rm ij}  }{\sum_{\rm kj\in{\cal I}} (A^{-1})_{\rm kj}}
 \sum_{\rm l\in{\cal B}} {\vec\pi}_{\rm l}\label{newpi}\\
 \nonumber\\
\langle {\vec \pi}_{\rm i}\cdot{\vec \pi}_{\rm j}\rangle& = & C^{\rm con}_{\rm ij} + \langle{\vec \pi}_{\rm i}\rangle \cdot \langle{\vec \pi}_{\rm j}\rangle  \\ \nonumber \\
C^{\rm con}_{\rm ij} &=&   2 \left [ A^{-1}_{\rm ij} -\frac{ \sum_{\rm kn\in{\cal I}} A^{-1}_{\rm ik}A^{-1}_{\rm nj}} {\sum_{\rm kn\in{\cal I}}A^{-1}_{\rm kn}}
\right ]
\label{corre_border2}
\end{eqnarray}
where we have separated the part of the correlation, $C^{\rm con}_{\rm ij}$, which is locally connected (i.e. the covariance).  

In Eq (\ref{newpi}) the first term describes a contribution coming from individuals on the border, while the second term is just a renormalization factor to ensure that $\sum_{\rm i \in{\cal I}}\langle\vec{\pi}_{\rm i}\rangle +\sum_{\rm l \in {\cal B}} \vec{\pi}_{\rm l}=0$. We can see from Eq (\ref{newpi}) that an individual $\rm i$ far from the border can also feel the effect of birds on the border, provided there is a nonzero connected correlation $C^{\rm con}_{\rm ij}$ between $\rm i$ and some individual $\rm j$  close to the border. In other terms, while {\em direct} mutual alignment occurs only between border individuals and immediate neighbors (for which ${\vec h}_{\rm i}$ are non zero), {\em effective} alignment  occurs also with internal birds that are indirectly correlated with them (for which $C^{\rm con}_{\rm ij}{\vec h}_{\rm j}$ are nonzero). If the connected correlations extend over sufficiently long distances, this mechanism ensures propagation of directional information trough the whole flock.

In Fig  \ref{fig-corre_border}B we show the behaviour of the connected correlation as a function of distance, for one flocking event. The scale over which this function decays, the correlation length, is of the order of the thickness of the flock (maximum distance between an internal point and the border), showing that $C^{\rm con}$ indeed  is long ranged enough to propagate ordering well inside the group. In the inset, we show that  the correlation length grows linearly with flock size, for all the flocking events we analyzed. Thus the correlation function is scale free: no matter how large the flock is, the correlation always extends over the whole flock.

\section{Correlation functions}
\label{App:correlations}

In this section we summarize the definitions of all the correlation functions introduced in the paper and we comment on their behaviour.

{\bf The pairwise correlation.}  Let us start by recalling the definition of the pairwise correlation,
\begin{equation}
\langle{\vec s}_{\rm i}\cdot{\vec s}_{\rm j}\rangle=\langle s^{\rm L}_{\rm i}s^{\rm L}_{\rm j}\rangle + \langle {\vec \pi}_{\rm i}\cdot {\vec \pi}_{\rm j}\rangle
\end{equation}
where, for future convenience, we have separated the longitudinal part of the correlation from the perpendicular one. We note that while the sample average of the perpendicular flight direction is zero, $(1/N)\sum_{\rm i}{\vec\pi}_{\rm i}=0$, the same is not true for the longitudinal direction. Rather, we have $(1/N)\sum_{\rm i}s^{\rm L}_{\rm i}=S$, and the longitudinal correlation is dominated by a contribution from the global polarization $S$. To  better investigate the degree of correlation in the system, it is then  convenient to focus on fluctuations of the individual flight directions with respect to the sample average. To this end, in all the figures in this paper we consider the following correlations, where we have subtracted the sample average contribution:
\begin{eqnarray}
C^{\rm P}_{\rm ij}&=&\langle {\vec \pi}_{\rm i} \cdot {\vec \pi}_{\rm j} \rangle \\
\nonumber\\
C^{\rm L}_{\rm ij}&=&\langle ({s}^{\rm L}_{\rm i}-S)(s^{\rm L}_{\rm j}-S)\rangle\nonumber\\
&=& \langle (1-S-\frac{1}{2}\langle \pi_{\rm i}^2\rangle)(1-S-\frac{1}{2}\langle \pi_{\rm j}^2)\rangle \label{long}\\
\nonumber \\
C_{\rm ij}&=&C^{\rm P}_{\rm ij}+C^{\rm L}_{\rm ij} \ .
\end{eqnarray}
The last identity in  Eq (\ref{long}) is a consequence of the spin wave approximation.

{\bf Connected correlations.} In Appendix \ref{App:fixedbound} we have described a theory where we get nonzero expectation values for the flight directions of individual birds,  $\langle\vec{\pi}_{\rm i}\rangle\ne 0$. In this case, it may be useful to look at correlation functions which are {\em locally} connected, i.e. that describe how the individual bird flight direction fluctuates with respect to its own average value and not---as in the previously defined correlations---with respect to the sample average. To this end, we have introduced in Appendix \ref{App:model} the following connected correlation function
\begin{equation}
C^{\rm con}_{\rm ij} =\langle {\vec \pi}_{\rm i}\cdot{\vec \pi}_{\rm j}\rangle- \langle{\vec \pi}_{\rm i}\rangle \cdot \langle{\vec \pi}_{\rm j}\rangle   \ .
\end{equation}
We note that in our case $C^{\rm con}_{\rm ij}$  is purely a theoretical construct. Indeed, we have applied the maximum entropy approach to each single snapshot independently. For a single snapshot, the experimental measurement of the correlation only consists in one configuration (the velocity field at that instant of time) and we cannot distinguish between connected and non--connected correlations. The only quantity that can be compared between theory and experiments is therefore $\langle \vec{\pi}_{\rm i}\cdot \vec{\pi}_{\rm j}\rangle$.

 \begin{figure}
\includegraphics[width = 0.6 \linewidth]{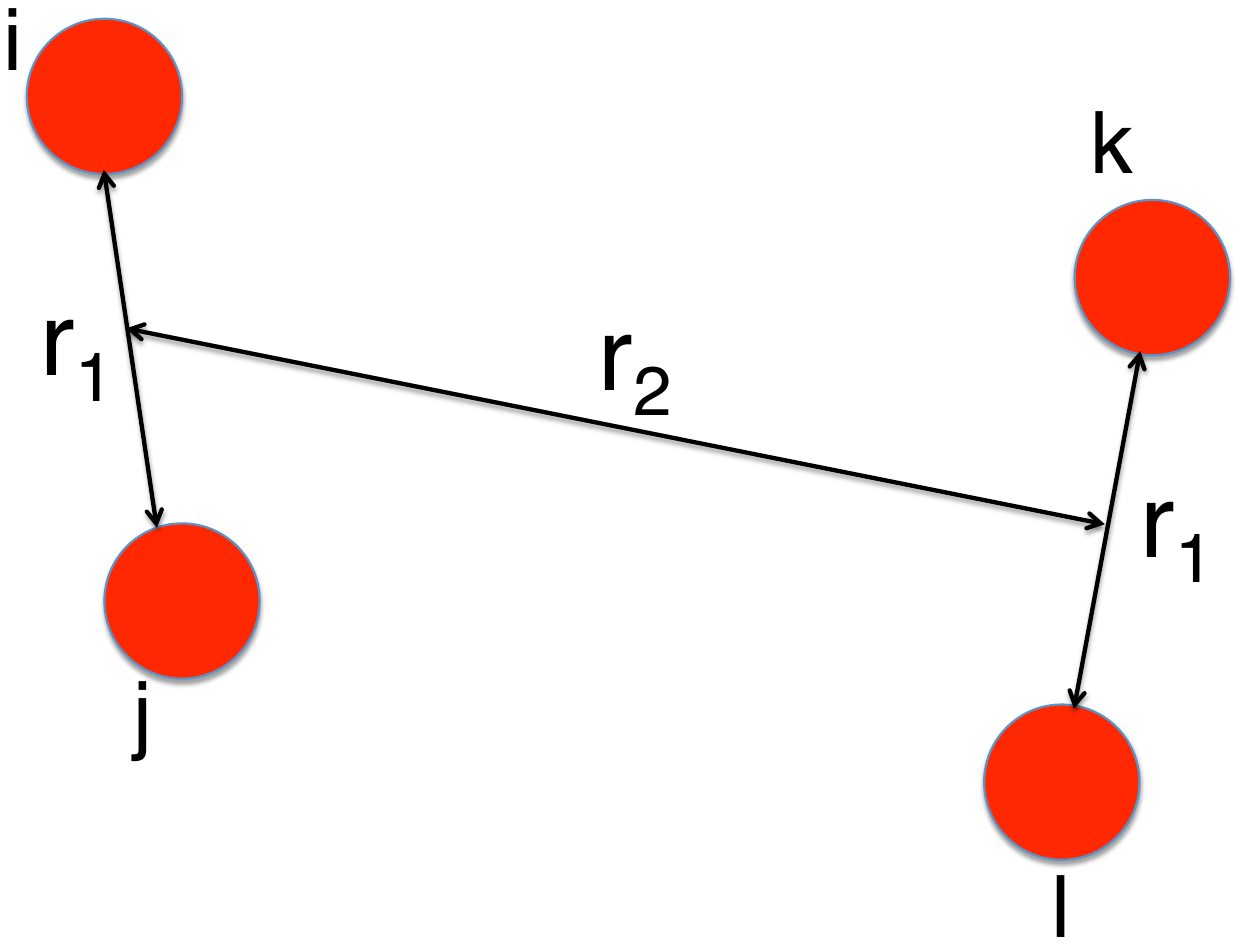}
\caption{Sketch of the structure of the four point correlation function. Red circles represent birds. Birds i and j have mutual distance $r_{\rm ij}=r_1$; birds k and l also have mutual distance $r_{\rm kl}=r_1$. The distance between the mid-point of the ij pair and the midpoint of the kl pair is $r_2$.}
 \label{corre4}
\end{figure}

{\bf  The degree of direct correlation.}  One important quantity entering our computation is the degree of direct correlation,
\begin{equation}
C_{\rm int}=\frac{1}{N}\sum_{\rm i}\frac{1}{n_c}\sum_{\rm j\in n_c^i} \langle{\vec s}_{\rm i}\cdot{\vec s}_{\rm j}\rangle ,
\end{equation}
which measures the average  correlation between an individual and its first $n_c$ neighbors. This degree of direct correlation is a single scalar quantity, and represents the input observable used by our maximum entropy approach to build a statistical model for the flight directions.

{\bf The two--point correlation function.}  To describe the behaviour of the correlation at different scales, it is convenient to define the two--point correlation function
\begin{equation}
C(r)= \frac{\sum_{\rm ij} C_{\rm ij}\delta(r_{\rm ij}-r)}{\sum_{\rm ij} \delta(r_{\rm ij}-r)} \ ,
\label{corre2}
\end{equation}
where $r_{\rm ij}=|{\bf\vec r}_{\rm i}-{\bf\vec r}_{\rm j}|$ is the distance between bird $\rm i$ and bird $\rm j$ and the delta function selects pairs of individuals that have mutual distances in a small interval around $r$ (the denominator representing the number of pairs in such an interval).  This function measures the average degree of correlation between individuals separated by a distance $r$.  Again it is possible to distinguish a  longitudinal and a perpendicular component of these correlations,
\begin{equation}
C(r)=C^{\rm L}(r)+C^{\rm P}(r) ,
\end{equation}
describing the contributions relative to, respectively, longitudinal and perpendicular fluctuations in the flight directions. Figure \ref{correlations} in the main text and Fig \ref{fig-corre_border} in the Appendix show the two--point correlation function computed from the maximum entropy model with fixed boundary conditions. The prediction  agrees  nicely with the experimental one, on all scales.  We stress that the maximum entropy model uses as an input only $C_{\rm int}$, which measures the average degree of correlation up to scale $n_c$. With the values of $n_c$ retrieved for our events ($n_c=21.2\pm 1.7$), this corresponds to a scale of the order of a few meters in $r$.  In contrast, the two--point correlation function measures the correlation on all possible scales, from close neighbors (a few meters) to the whole extension of the flock (hundreds of meters, for some flocks). Therefore, the good agreement with experiments represents a highly nontrivial prediction of the model.  From Eq (\ref{corre2}), the correlation function takes into account the contribution coming from all pairs of individuals, independent of whether they reside on the border or in the bulk of the flock. Yet, when adopting fixed flight directions on the border of the flock, the contribution coming from birds on the border is by construction identical in the predicted and  observed correlation functions. To test more explicitly whether the model provides good predictions for the correlations of internal individuals, we can consider an {\it internal} correlation function, defined as in Eq (\ref{corre2}), but where we only count contributions from individuals inside the flock (${\rm i}, {\rm j} \in {\cal I}$);  the result is in Fig \ref{fig-corre_border}A.  Again, the prediction of the model is nicely consistent with the experimental correlation.

{\bf The four--point correlation function.}   We can define correlation functions not only between pairs of individuals, but for more complicated arrangements of birds. For example, let us consider a pair of birds ${\rm i}, {\rm j}$ separated by a distance $r_1$, and measure their mutual alignment.  We might want to compare this degree of alignment to the one that another pair of birds ${\rm k}, {\rm l}$, also separated from one another by a distance $r_1$, that are located in another position in the flock. We can then define the following four--point correlation
\begin{eqnarray}
C_{4}(r_1;r_2)&=& \frac {
\sum_{\rm ijkl} \langle ({\vec \pi}_{\rm i}\cdot {\vec \pi}_{\rm j} )({\vec \pi}_{\rm i}\cdot {\vec \pi}_{\rm j})\rangle \Delta_{\rm ijkl}}
{ \sum_{\rm ijkl} \Delta_{\rm ijkl} },\\
\Delta_{\rm ijkl} &= &\delta(r_{\rm ij}-r_1)\delta(r_{\rm kl}-r_1)\delta(r_{\rm ij-kl}-r_2)   \nonumber\\
&&
\end{eqnarray}
where $r_{\rm ij-kl}$ indicates the distance between the midpoints of the pairs $\rm ij$ and pair $\rm kl$;  see Fig \ref{corre4}. We can plot $C_{4}(r_1;r_2)$  as a function of the two distances $r_1$ and $r_2$. For example, in Fig \ref{correlations}C in the main text, it is shown for a fixed value of $r_1$ as a function of $r_2$.  We also note that, in the spin wave approximation, the longitudinal correlation $C^{\rm L}$ is nothing else than a particular case of the four--point correlation, 
 \begin{equation}
 C^{\rm L}(r)=1-C_{4}(0;r) -S^2 .
 \end{equation}

\end{document}